\documentclass{article}
\usepackage[utf8]{inputenc}
\usepackage[margin=1.2in]{geometry}
\usepackage{color, url}
\usepackage{graphicx}
\usepackage{braket}
\usepackage{amsmath}
\usepackage{float}
\usepackage{bbding}
\usepackage{hyperref}
\usepackage{makecell}
\usepackage{hyphenat}
\usepackage{lineno}
\usepackage{hhline}
\usepackage{multicol,lipsum}
\usepackage{makecell}
\usepackage{algorithm}
\usepackage{array}
\usepackage{booktabs}
\usepackage{soul}
\usepackage{babel}
\usepackage{subcaption}
\usepackage{hyperref}
\usepackage{adjustbox}
\usepackage[autolanguage]{numprint}
\usepackage[]{footmisc}
\usepackage{bbding}
\usepackage{multirow}
\usepackage{amssymb}
\usepackage{latexsym}
\usepackage{authblk}
\usepackage{comment}
\usepackage{url}
\usepackage{xcolor}
\definecolor{newcolor}{rgb}{.8,.349,.1}

\hypersetup{
    colorlinks=true,
    linkcolor=blue,
    filecolor=magenta,      
    urlcolor=cyan,
}

\title{Physical Realization of Measurement Based Quantum Computation }

\author{Muhammad Kashif \\  \href{mailto:}{mkashif@hbku.edu.qa} 
   \and Saif Al-Kuwari \\  \href{mailto:}{smalkuwari@hbku.edu.qa} }
\affil[]{\textit{Division of Information and Computing Technology, College of Science and Engineering, Hamad Bin Khalifa University, Qatar Foundation, Doha Qatar}}
\date{}

\begin{document}
\maketitle

\begin{abstract}
Harnessing  quantum mechanics properties, quantum computers have the potential to out-perform classical computers in many applications and  are envisioned to affect  various  aspects of our society. Different  approaches  are  being  explored  for  building  such  computers. One of such potential approaches is Measurement based quantum computation (MBQC), introduced by Raussendorf and Briegel in 2001. In MBQC a large number of qubits are prepared in a highly entangled clusters, called cluster states. The required quantum computation is then performed by a sequence of measurements.  Cluster states are being physically realized using continuous variables (CV) and discrete variables (DV) approaches. CV-based approaches can be further categorized as Frequency domain multiplexing (FDM), Time domain multiplexing (TDM), Spatial domain multiplexing (SDM) and hybrid. We discuss and compare these approaches in detail. We also discuss cluster states generation in DV and report some recent results where photons and superconducting qubits are used.    
\end{abstract}
\paragraph{Keywords.}
   Continuous variables cluster states, Discrete variables cluster states, Measurement based quantum computation, One-way quantum computation,  Physical realization, Quantum computation.

\section{Introduction}
\label{sec:introduction}
The race for computational capabilities began during the second world war with Alan Turing’s Universal Turing Machine. It was the first general purpose digital computer with an electronically stored program \cite{Copeland:revised2020}. In 1945 , Von Neumann redesigned Turing’s idea with some modifications, which is now the most popular computer architecture used by almost all current computers, efficiently solving a wide range of significant problems \cite{Marella:2020}. Since the size of electronic devices for conventional computer is rapidly shrinking, quantum effects during the fabrication process are starting to disrupt the functionality of these devices, potentially predicting the end of Moore’s law \cite{nielsen:2002}, which states that computational power doubles around every 18 months. Moreover, there are some problems which cannot be solved by these classical computers in practical amount of time \cite{Marella:2020}, even if Moore's law continue to sustain. Therefore, a new approach to perform computation is necessary. One proposal is Quantum Computation.

In 1980s, Richard Feynman was among the first to promote the idea that quantum computer can efficiently solve some computationally intensive problems in physics and chemistry \cite{Feynman:1982}. Quantum computing is a promising new approach of computation, which can harness the properties of quantum mechanics \cite{Marella:2020}, and has the ability to perform certain tasks with exponential speed compared to classical computers \cite{Arute:2019}. In 1994’s, Peter Shor discovered a quantum algorithm that can efficiently solve mathematical problems to the core of modern cryptography, potentially breaking public-key cryptography \cite{Shor:1994}. Similarly, in 1996 Lov Grover proposed a quantum search algorithm \cite{Grover:1996} that can search through unstructured data in polynomial time (more efficient that any classical computer can achieve). Other quantum algorithms \cite{bernstein:1997,Deutsch:1992, QuantumAlgorithmZoo}, further proved this quantum advantage and showed that no classical algorithms can provide similar performance. Since then, the quest for developing a quantum computer began and a significant progress has been made till date.
Recently, a practical proof of quantum advantage has been demonstrated \cite{Arute:2019,zhong:2020, wu:2021} through the so-called Quantum Supremacy, where a real quantum computer was able to perform a task in seconds that would otherwise require a super (classical) computer thousands of years to run.

The fundamental difference between classical computers and quantum computers is the concept of state \cite{Mcgeoch:2014}. Unlike classical bits, quantum computers use quantum bits also known as qubits, for storing and manipulating quantum information. Any two-level quantum system like spin and  polarization, can form a qubit \cite{OLIVEIRA2007}.

Unlike the bit that can be either in 0 or 1 state, a qubit can be in combination of both states at the same time until it is observed (measured), which forces it to collapse to either 0 or 1, such phenomena is known as superposition \cite{nielsen:2002}. Along with superposition, quantum entanglement is another important quantum phenomenon and has no classical counterpart. When two qubits are entangled their states become tightly correlated such that the measurement of one quantum object inevitably affects the measurement of the other \cite{Marella:2020} \cite{Jozsa:1997}.
In fact, superposition and entanglement are the two fundamental properties behind the computational power of quantum computers \cite{nielsen:2002}. 

\subsection{Types of Quantum Computation} 
Quantum computation can be broadly divided into two categories: Adiabatic quantum computation (quantum annealing) and Universal quantum computation. Adiabetic quantum computation does not use qubit gates, instead analog values are manipulated in Hamiltonian \cite{Marella:2020}. Hamiltonian of a system in quantum mechanics is an operator representing the total energy of the system (potential and kinetic energy). On the other hand universal quantum computation can be achieved through various approaches, however, the most prominent is gate-based quantum computation.\footnote{Adiabatic and gate-based quantum computation have shown to be equivalent but are not the same class of quantum computation}

\paragraph{Adiabatic Quantum Computation.} Adiabatic quantum computation is an analog quantum computation technique where qubits with initial quantum states are changed to Hamiltonian in a way that the problem to be solved is encoded in the final Hamiltonian, which corresponds to the output \cite{albash:2018}. The adiabatic quantum computation refers to a phenomenon when the system remains in the ground state of changing Hamiltonian \cite{Mcgeoch:2014}. Adiabatic quantum computation exploits quantum annealing for solving optimization problems for various applications including machine learning \cite{lloyd:2013,Benedetti:2017,Li:2018}, classification tasks \cite{neven:2009}, variational auto-encoders \cite{Khoshaman:2018,wilson:2019} and compressive sensing \cite{ayanzadeh:2019}. Every qubit state can be represented as an energy level in quantum annealers. These states are simulated for a given application and the lowest energy results are obtained. The optimal solution is provided by a state with lowest energy. Adiabatic quantum computers are among the first commercially available quantum computers, mainly popularized  by D-wave Systems \cite{johnson:2011}, which produce programmable quantum annealers.

\paragraph{Universal Quantum Computation.} Universal quantum computers can complete tasks that are beyond the reach of current classical computers \cite{Marella:2020}. Gate-based quantum computation is one of the most-widely used approaches for universal quantum computation and uses quantum gates to perform operations on qubits. This approach often use the quantum circuit model \cite{Marella:2020} to perform a sequence of operations and measurements on qubits. IBM and google (and many other industry giants and startups) have already built gate-based quantum computer with less than 100 qubits \cite{Mcgeoch:2014, hauke:2020}, and limited amount (typically 5-10) of qubits to use for simulation purposes \cite{Marella:2020}.
Building universal quantum computers with sufficiently large number of qubits, typically thousands of qubits, is quite challenging, but when built they are anticipated to solve highly computationally intensive tasks within practical amount of time.



To build a universal quantum computer, one must be able to generate reversible and deterministic entanglement. Therefore, initially the community was mainly focusing on entangling the closely located qubits, typically nanometers apart \cite{kane:1998, Loss:1998}. However, locating qubits that are close to each other, without losing control of individual qubits, is both challenging and unscalable \cite{Benjamin:2009}. To overcome this problem, an alternative approach was proposed by Raussendorf and Briegel, in 2001, called: measurement-based quantum computation (MBQC) \cite{Raussendorf:2001}, which is also commonly called  one-way quantum computer (1WQC). MBQC performs the quantum computations only by exploiting single qubit measurements in different measurement bases via a highly entangled resource state of qubits. We discuss more details about MBQC in section \ref{MBQCsection}.

Quantum gates with high fidelity rates \footnote{Fidelity rate is measure of similarity between two quantum states\cite{fidelityOnline:2019}} are key to efficient quantum information processing (QIP) \cite{nielsen:2002}, which are directly affected by control imperfections and decoherence \footnote{Decoherence is the loss of quantum coherence that makes the qubit entangled with the environment resulting in an inaccurate quantum computation\cite{shor:1995}}.
Upgrading the fabrication processes can mitigate the decoherence \cite{Joos:2009}, but this requires extensive research at the material that is being used to realize qubits \cite{Krantz:2019,Wendin:2017}. Control imperfections, including signal distortion and instrumental instability, can be mitigated by using exquisite and complex calibration processes \cite{Arute:2019}. The fidelity rate of quantum gates is usually measured via randomized benchmarking \cite{Magesan:2011}.

Functional quantum computers, most notably based on superconducting qubits, have already been developed(though with limited number of qubits, $<100$), in order to fully utilize the computational power of quantum computation, a significantly large (tens of thousands) number of qubits are required \cite{Cirac:2000}, and because of the above-mentioned limitations scaling up beyond  “quantum supremacy” is experimentally challenging in existing quantum computers. This is where MBQC/1WQC may prove an advantage. 
Cluster or graph states, which are building blocks of 1WQC \cite{Chen:2007}, have shown entanglement robustness against decoherence \cite{Hein:2005}.
Hence the 1WQC, sometimes also referred as cluster state model, is well suited for scalable quantum computation and advances in research have already been made for physical realization of cluster states \cite{Yang:2005,Vallone:2008,browne:2006,Fulconis:2007,Lee:2012}.

\subsection{Related work.} 
Earlier proposals on MBQC were mostly focusing on  general MBQC, its universality proof and its potential in achieving scalable quantum computation model. In recent years we observe a shift toward the experimental realization of cluster states to demonstrate that MBQC is indeed one of the leading candidates towards scalable Quantum computation. 
There are various review papers on physical realization of quantum computers \cite{resch:2019,Suominen:2012,Jazaeri:2019,xin:2018}. However, despite its potential,  there is no dedicated work in the literature surveying the state-of-the-art physical realizations of MBQC except \cite{Pfister:2019}, which only reviews one of subtypes of MBQC's physical realization techniques (Frequency domain multiplexing) for continuous variables cluster states. 
A recent review on general MBQC is published in \cite{Tzu:2021}, with no focus on physical realization. A brief scope comparison of our survey with some recent related surveys is presented in Table \ref{tab:contribution_comparison}.  Hence, a detailed review of the approaches being used for implementing cluster states for MBQC, is the ultimate need of time providing the research community a sound overview of what has been achieved in this regard and what potentially can be achieved.  
\begin{table}[!t]
    \centering
    \caption{Scope comparison of this survey with recent surveys on MBQC}
    \begin{tabular}{p{1cm}|p{1cm}|p{1cm}|p{1cm}|p{1cm}|p{3cm}}
    \toprule
         \multirow{2}{*}{Ref\#}  &\multicolumn{4}{c|}{\makecell{CV physical realization\\ of cluster states}} &\multirow{2}{*}{\makecell{DV physical realization\\ of cluster states}} \\
        \cline{2-5}
         &FDM &TDM &SDM &Hybrid &\\
        \bottomrule
         \cite{Pfister:2019} &$\hfil\checkmark$ &$\hfil\times$ &$\hfil\times$ &$\hfil\times$ &$\hfil\times$\\
         \hline
         \cite{Tzu:2021} &$\hfil\times$ &$\hfil\times$ &$\hfil\times$ &$\hfil\times$ &$\hfil\times$ \\
         \hline
         Ours  &$\hfil\checkmark$ &$\hfil\checkmark$ &$\hfil\checkmark$ &$\hfil\checkmark$ &$\hfil\checkmark$\\
         
    \bottomrule
    \end{tabular}
    
    \label{tab:contribution_comparison}
\end{table}

\subsection{Contribution and Scope.} 
In this paper, after providing a somewhat detailed discussion on MBQC and its universality, we focus on the techniques and approaches proposed for physical realization of MBQC. In general, two main approaches are discussed: Continuous Variables (CV) and Discrete Variables (DV) as being used for cluster states realization. We further categorize CV based approaches into four subcategories: frequency domain multiplexing, time domain multiplexing, spatial domain multiplexing and hybrid (combination of any two multiplexing approaches). We also provide a comparison on the recent state-of-the-art proposals on all these techniques. Moreover, the state-of-the-art cluster state realization in DV, particularly in photonic qubits are summarized and compared. In addition, we also survey and compared the recent advances in DV for qubits entanglement, specifically in photonic and superconducting qubits, which can potentially be used as a resource state for MBQC. 

\subsection{Organization} 
The rest of this paper is organized as follows: Section \ref{MBQCsection} provides an overview of MBQC and the role of cluster states in MBQC. Universality of MBQC is discussed in this section too. Section \ref{MBQC_PR} provides a general overview of the main approaches of physically realizing MBQC. This is followed by Sections \ref{CVQC} and \ref{DVQC}, where we discuss in detail the main approaches adopted for physical realization of MBQC, namely the CV-based and DV-based quantum, respectively. These sections also report on the existing experimental results of such approaches. Finally, the paper concludes in Section \ref{conclusion}, where we provide some concluding remarks and point out some promising directions toward realizing a practical a quantum computer based on MBQC. 



\section{Measurement Based Quantum Computation (MBQC)} \label{MBQCsection}

\begin{figure}[!htp]
	\centering
		\includegraphics[scale=0.5]{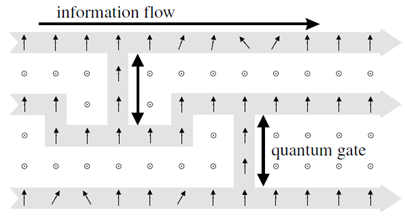}
	\caption{Quantum Information Processing in MBQC \cite{Raussendorf:2001}. The tilted arrows represent measurement in x-y plane, vertical arrows represent measurements in the eigen basis of $\sigma^x$ and circle represent measurements in the eigen basis of $\sigma^y$.}
	 \label{fig:MBQC}
\end{figure}

In MBQC all the computations are performed via measurements exploiting a highly entangled resource state (also called cluster state). The two well-known schemes for MBQC are one-way quantum computation (1WQC) model and teleportation-based model. In one-way quantum computation, single qubit measurements are used whereas joint measurements (entangled measurements) are used in teleportation-based model to achieve universal quantum computation \cite{briegel:2009}. However, in teleportation-based model, all the qubits present in the system are measured separately in a specific order and measurement basis. The actual algorithm is specified by this measurement order. On the other hand, in one-way quantum computation, generation of entanglement is no longer part of the quantum algorithm being developed. In MBQC, the execution of an algorithm is done via the measurements only, by exploiting an already generated entangled state \cite{Benjamin:2009,nielsen:2006}. The system is prepared in a highly entangled quantum state known as cluster state with no dependence on the quantum algorithm being developed \cite{briegel:2009}. This study focuses on 1WQC scheme of MBQC\footnote{In this paper, 1WQC and MBQC are used interchangeably in this paper}. However, readers interested about teleportation-based model are referred to \cite{jozsa:2006}. A typical workflow of 1WQC begins by creating such cluster state, encoding the information onto the cluster state and finally perform information read-out via single-qubit measurements \cite{browne:2006}. It is also called \emph{one-way quantum computer} because the resource state entanglement can only be measured once, after which it will be destroyed \cite{raussendorf:2003}. 
Cluster states can be created by using quantum-Ising type interaction in lattice configuration between two-state particles. More details on cluster states are in Section \ref{clusterstates}. A typical representation of MBQC, as in original paper \cite{Raussendorf:2001}, is shown in Fig. \ref{fig:MBQC}.

When it comes to the development of universal quantum computer, 1WQC, promise comparable  potential to that of the circuit-based model which is reversible in nature \cite{raussendorf:2012}.
In MBQC, quantum information is processed by performing a sequence of adaptive measurements \cite{briegel:2009}. For details on how adaptive measurements are performed, refer to \cite{jozsa:2006}. 
MBQC offers a number of advantages that makes it a potential approach to build  real scalable quantum computer.
\begin{itemize}
    \item In MBQC, typically one-way quantum computation, an entangled state is generated separately well in advance, which makes it fault-tolerant, as errors can be recognized without harming the algorithm being implemented \cite{briegel:2009}. 
    
    \item In 1WQC scheme of MBQC, the entire computation resource is provided by specific entangled state, allowing to track the computations all the way back to the entangled resource state, which may help in maximizing the computational advantage of quantum computation\cite{Raussendorf:2001}.
    
    \item MBQC only requires nearest-neighbor Ising coupling rather than tunable interactions between qubits which makes it more scalable and parallelized \cite{raussendorf:2003}
    
    \item An appropriate combination of MBQC and topological error correction provides strong basis towards noise resilient scalable quantum computer \cite{briegel:2009}. Details of error correction is not within the scope of this review.
\end{itemize}

On the other hand, compared to the circuit model, MBQC requires a relatively large number of qubits to be stored at the same time. However, ‘on the fly’ (Section \ref{CV_exp_realization}) creation of cluster states can help in overcoming this issue.

\subsection{Cluster states} \label{clusterstates}
As computation in MBQC is dependent on cluster state, it is vital to shed some light on them. The term cluster state, originally penned by Raussendorf and Briegel \cite{briegel:2001}, is a group of highly entangled quantum states, exhibiting two very prominent characteristics: entanglement persistency and maximal connectedness of the entangled state.  Entanglement persistency is the ratio of amount of entanglement in multi-particle system to operational effort for destruction of all the entanglement in the system.  Maximal connectedness means that by using single-qubit measurements, every qubit pair can be projected with certainty to maximally entangled state. Later on, as described in section (\ref{MBQCsection}), it was shown in \cite{Raussendorf:2001}, that 1WQC can be built using these cluster states by exploiting single-qubit measurements only. 

Since the entanglement cannot be added to the system by single qubit operations, it can be inferred that in order to generate any type of entanglement from the resource state, it is essential that the resource cluster state already contains that entanglement, which can be measured using an appropriate entanglement measurement \cite{Girard:2017}. This means that maximally entangled states is what makes a resource and/or cluster state universal \cite{Nest:2006}\cite{Nest:2007}, which helps to determine the potential cluster states for MBQC. Moreover, these cluster states should be multi-partite entangled states.

Although states like Dicke states, certain ground states of strongly correlated 1D spin systems, W-states and and $\lambda$-particle 1D cluster states are considered highly entangled states, they are not universal in this regard \cite{Nest:2006} because of the fact that there exist at least one non-maximal type of entanglement is these states. However, there are some states which can be universal cluster states for MBQC since they fulfill all the entanglement criteria. These states include graph states \cite{Hein:2006}, which are combined with triangular, hexagonal, Kagome kinds of regular 2D lattices. Also, the lattices with high degree of defects can serve as universal cluster state for MBQC \cite{Nest:2006}. 
Cluster states can be visualized as a graph, sometimes called \emph{graph states}. Graph states can be defined as graph with a set of vertices and edges connect the pair of vertices \cite{raussendorf:2003,Hein:2006}. Cluster states are subclass of graph states with an $n$-dimensional square grid as an underlying graph \cite{browne:2006}. Preforming computation on cluster state proceeds in three steps \cite{nielsen:2006}: (1) Preparation of cluster state, which is highly entangled many-qubit state; (2) Processing the cluster via an adaptive sequence of single-qubit measurements; (3) Using the remaining qubits for result read-out of the underlying computation. Below, we will discuss procedures for cluster states generation and preparation, measurement processing and output in cluster states. 

\begin{figure}[!htp]
    \centering
    \begin{subfigure}[b]{0.2\textwidth}
        \centering
        \includegraphics[scale=0.2]{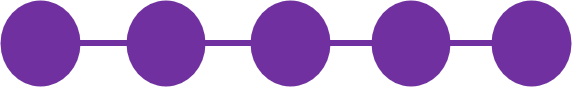}
        \caption{}
    \end{subfigure}
    \hspace{1cm}
    \begin{subfigure}[b]{0.2\textwidth}
        \centering
        \includegraphics[scale=0.2]{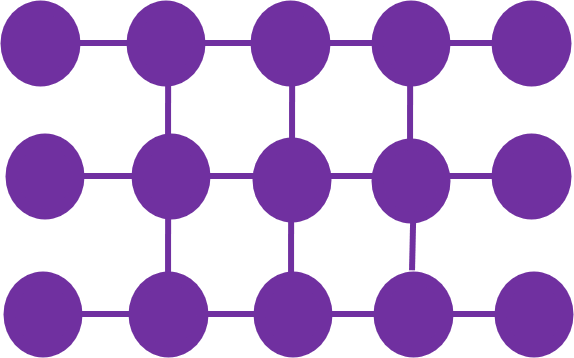}
        \caption{}
    \end{subfigure}
    \caption{Cluster States Generation and Preparation (a)1D cluster state. The circles represent qubit states and the line between two circles represents the controlled-phase operation between the corresponding qubits. (b) 2D cluster state
  }
\label{fig:clusterstatefig}
\end{figure}

\paragraph{Cluster State Generation and Preparation.} \label{cluster_prep}
In context of MBQC, an $n$ qubit cluster state can be generated by exploiting any two-dimensional (2D) and/or one-dimensional (1D) lattice graph with $n$ number of vertices \cite{nielsen:2006}, as shown in Fig. \ref{fig:clusterstatefig}, where a single qubit is assigned to each vertex. Once generated, a possible approach for cluster state preparation is presented in \cite{nielsen:2002}, according to which the following two steps can be performed for cluster state preparation: (1) Preparing all the qubits in superposition state, i.e., $\ket{+}= \frac{1}{\sqrt{2}}(\ket{0} + \ket{1})$; (2) Apply controlled-phase gates between the two connected qubits. The order of these gates is irrelevant since these gates commute\cite{raussendorf:2012}




 
\paragraph{Measurement Processing and Output in Cluster States.}
The next step after preparing the cluster state is performing a sequence of processing measurements with following attributes\cite{Dawson:2006}: (1) All measurements are single qubit measurements; (2) Previous measurement results may help in selecting the measurement basis – allowed to exploit feedforward of classical measurement (3) A classical computer can process these measurement results. 

Once processing is completed, the outcome of the computation can be returned in two ways \cite{nielsen:2006,Dawson:2006}: (1) Quantum state $\ket{\psi}$ as outcome: once the processing measurement sequence is completed, output the state of remaining qubits (not measured); (2) Adding a set of read-out measurements: single-qubit measurement sequence is applied to the remaining qubits after the completion of all processing (result is a classical bit string).

\subsection{Stabilizer Formalism} \label{stabilizer_formalism}
The number of parameters required to describe quantum states increases exponentially with the number of qubits, making the underlying quantum system more complex and difficult to understand\cite{Fujii:2015}. This problem motivate investigating techniques that can help understanding these complex quantum systems. The stabilizer formalism \cite{Gottesman:1997}, is one such technique which helps in understanding the complex quantum systems and underlying operations primarily because of the compact description and characterization of quantum states and sub-spaces over multiple qubits, and their evolution under Pauli measurements\footnote{Pauli measurements correspond to Pauli matrices ($X$, $Y$ and $Z$) in quantum computation.} and Clifford group \footnote{A collection of unitary operators that map Pauli group (a group comprising tensor product of Pauli matrices $n$ times)  onto itself is known as Clifford group for $n$ qubits. }\cite{chung:2009}.

Instead of components (in some basis), of the state itself, the stabilizer formalism specifies a set of eigenvalue relations to describe a state of a quantum system. A stabilizer can then be defined as follows\cite{browne:2006}: An operator $K$ is said to be stabilizer for a sub-space $S$ when $K\ket{\psi} = \ket{\psi}$ for all $\ket{\psi}\in S$, that is $\ket{\psi}$ is an eigenstate of $K$ with eigenvalue +1. It is important to note here that joint eignenstates with eigenvalues of -1 also exists, however by definition, only the eigenstates with eigenvalue of +1 are stabilized. 

The objective in stabilizer formalism is to come up with a set of operators which not only have the stabilizing property mentioned above, but are also Hermitian members of Pauli group.  Identification of stabilizing operators which uniquely defines the given state or sub-space (no state outside the sub-space), is key in stabilizer formalism, and these uniquely defined states\slash sub-spaces are called stabilizers states\slash subspaces \cite{browne:2006}. These operators exhibit all properties of the states allowing an easier analysis of how a given state changes under unitary evolution and measurements. The set of operators stabilizing a sub-space has a group structure (also known as stabilizer group), because the product of two stabilizing operators is itself stabilizing. In quantum information processing, stabilizer states and subspaces occur in many states including GHZ states, Bell states, different error correcting codes, cluster states and graph states. 

The stabilizer formalism was originally proposed for describing quantum error correction codes but it is also applicable to MBQC. Conventionally, $2^n$ complex number are required for a complete description of an $n$-qubit state. However, a group of $n$ stabilizers can describe an $n$-qubit state, which is stabilized by Pauli group \cite{Gottesman:1997}. In other words, an entire set of commuting observables can be created by $n$ stabilizers, resulting in a more compact description of a quantum system. Moreover, the stabilizer formalism can efficiently describe the underlying unitary evolution of Clifford Group operators. More importantly, the stabilizer formalism has proven to be quite efficient for state description of a multi-qubit system under projective measurements in $X$, $Y$ and $Z$ basis \cite{chung:2009}.

\subsection{Universality of MBQC} \label{MBQC_universality}
In order to perform arbitrary quantum computations, single and multi-qubit quantum gates are required. Any quantum approach that can realize these quantum gates can be considered a universal quantum computation approach. The one-way quantum computation scheme of MBQC can perform any quantum computation, and has been proven to be universal, with various single and multi-qubit gates realization \cite{raussendorf:2003}. Below we discuss the realization procedures of H-gate, $\pi/2$-phase gate and CNOT gate on MBQC, which, together, can form form any quantum gate.

\begin{figure}[!htp]
	\centering
     \begin{subfigure}[b]{0.2\textwidth}
         \centering
         \includegraphics[scale=0.25]{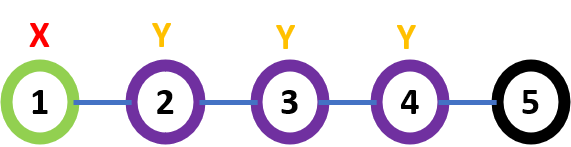}
         \caption{}
          \label{fig:H_gate_mbqc}
     \end{subfigure}
     \hspace{1cm}
     \begin{subfigure}[b]{0.2\textwidth}
         \centering
         \includegraphics[scale=0.25]{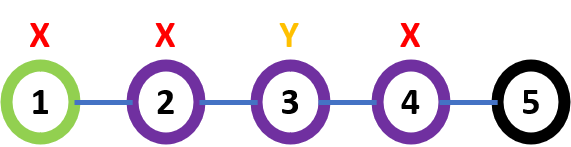}
         \caption{}
          \label{fig:pi/2_gate_mbqc}
     \end{subfigure}
     \hspace{1cm}
        \begin{subfigure}[b]{0.3\textwidth}
         \centering
         \includegraphics[scale=0.25]{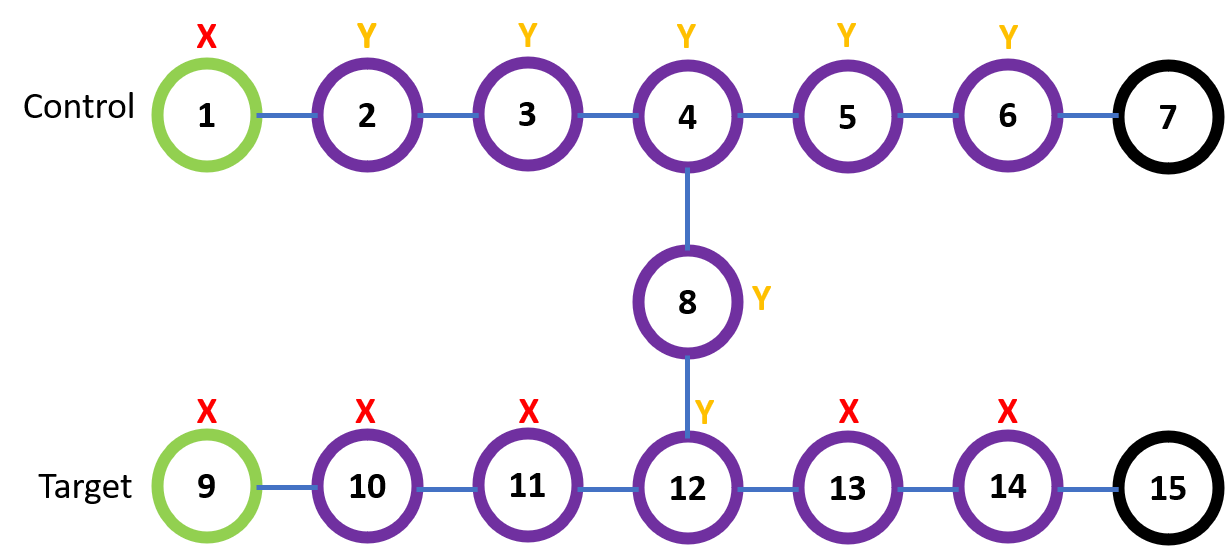}
         \caption{}
          \label{fig:CNOT_gate_mbqc}
     \end{subfigure}
     \hfill
	\caption{Graph state representation of H $\pi/2$ and CNOT-Gate. The circles represents the qubits, the blue lines represent the CZ operation, the green and black circles indicates the input and output qubits, respectively. The symbols $X$ and $Y$ represent the measurement-bases for the corresponding qubit measurement. (a) H-gate representation; (b) $\pi/2$-phase gate realization using MBQC; (c) CNOT representation}
	\label{fig:H-pi-2-gate_cluster}
\end{figure}

\paragraph{Hadamard Gate Realization.}
Hadamard gate is an important single qubit gate in quantum Computation responsible for putting a given qubit state in superposition. In MBQC realm, the H-gate can be realized  by performing these steps (the corresponding graph representation as presented by \cite{raussendorf:2003} is shown in Fig. \ref{fig:H_gate_mbqc}): (1) Prepare all qubits in the $\ket{+}$ state except the first qubit ( the input qubit on  Fig. \ref{fig:H_gate_mbqc}); (2) Entangle the connected qubits using CZ gate; (3) Measure the first Qubit in eigen basis of $\sigma^x$ and measure qubits 2-4 in eigen basis of $\sigma^y$. 

Once the sequence of measurements being applied on qubits 1-4, the output qubit from Fig.  \ref{fig:H_gate_mbqc} goes to superposition state. 

\paragraph{\texorpdfstring{$\pi$/2-Phase Gate Realization.}{}}
Another important single qubit gate is $\pi/2$-phase gate. Similar to H-gate, $\pi/2$-phase gate can also be realized on MBQC with five qubits in linear configuration, but with slightly different qubit measurement basis. Fig. \ref{fig:pi/2_gate_mbqc} depicts the corresponding cluster state. $\pi/2$-phase gate is prepared as follows: (1) Prepare all qubits in the $\ket{+}$ state except the first Qubit (the input qubit on Fig. \ref{fig:pi/2_gate_mbqc}); (2) Entangle the connected qubits using CZ gate; (3) Measure qubits 1,2 and 4 in eigen basis of $\sigma^x$ and measure the third qubit in eigen basis of $\sigma^y$. 
After the sequence of measurements being applied on qubits 1-4, the phase of output qubit from Fig. \ref{fig:pi/2_gate_mbqc} is rotated $\pi/2$ times.

\paragraph{CNOT Gate Realization.}
 
The CNOT operation can be realized in MBQC with 2D cluster state as shown in Fig. \ref{fig:CNOT_gate_mbqc}. The steps required for CNOT gate realization are as follows: (1) Prepare all qubits in $\ket{+}$ state except input qubits (1 and 9) from Fig. \ref{fig:CNOT_gate_mbqc}; (2) Entangle all the connected qubits from Fig. \ref{fig:CNOT_gate_mbqc}  using CZ gate; (3) The output qubits (7 and 15) are not measured. The qubits 1,9,10,11,13 and 14 are measured in eigen basis of $\sigma^x$ and all remaining qubits in eigen basis of $\sigma^y$.   



\section{Physical Realization of MBQC} \label{MBQC_PR}
Measurements and entanglement are the core properties MBQC relies on. In fact, the 1WQC completely relies on single-qubit measurements and entanglement between qubits, as discussed in Section \ref{MBQCsection}. Entanglement is provided well in advance via a highly entangled cluster state, which is a central resource of quantum information processing.  Hence, the physical realization of MBQC is highly dependent on how the cluster states are being physically realized. In 1WQC, adaptive single-qubit measurements are performed to implement a certain algorithm with feed-forward operations governed by previous qubit measurement outcomes \cite{Raussendorf:2001}. Two different feed-forward operations can be considered: previous measurement outcomes direct the selection of new measurement basis and the feed-forward corrections based on Pauli matrix on the output state \cite{Vallone:2008}.

1WQC is originally proposed by Raussendorf and Briegel for discrete variables \cite{Raussendorf:2001}, and is later extended to CV domain by Menicucci \cite{Menicucci:2006}. In DV, observables are usually realized on single-photon states known as photonic qubits \cite{Su:2012}, while in CV observables are usually realized on optical modes also known as qumodes  \cite{Braunstein:2005}.
The degree of scalability for an effective QIP is the main obstacle towards the physical realization of cluster states. The scalability, at material level, deals with the limitations on the topology and size of cluster states \cite{Korolev:2018}. This limitation directly restricts the number of logical operations needed to process large volumes of data. The nature of such limitations varies depending on the materials used to physically realize qubits in a cluster state. As mentioned in Section \ref{sec:introduction}, the physical realization approaches for cluster states realization in MBQC context can be broadly divided into two classes: continuous variables \cite{Mile:2009,Larsen:2019,Peter:2014,Kui:2014,Mamaev:2019} and discrete variables \cite{Korolev:2018,Zhou:2003,Schwartz:2016} cluster states.

The structure of cluster state of qubits (for DV based quantum computation) or qumodes (for CV-based quantum computation), determines the computational characterstics of MBQC \cite{Asavanant:2019}. One-dimensional (1D) cluster states are capable of single qubit or single qumode operations. However, two-dimensional (2D) cluster states are key for universal quantum computation. Unlike 1D, where qubits\slash qumodes are entangled as a single chain, in 2D cluster state,qubits\slash qumodes are entangled in a 2D lattice configuration. The number of qubits or qumodes determines the cluster state scalability or the number of operations. Consequently, for universal MBQC, realization of a large-scale two-dimensional cluster state is important. 2D qubit-cluster states, despite being proposed on various physical systems \cite{Yokoyama:2013,Yoshikawa:2016,yang:2016,Alexander:2016,Grimsmo:2017,Albarr:2018,Mamaev:2019}, is quite challenging to implement experimentally. For instance, in case of stationary qubits \cite{Albarr:2018,Mamaev:2019}, such as ion traps and superconducting qubits, a large number of qubits are required to be prepared and spatially arranged for a large-scale cluster state realization. As a result, the experimental complexity increases directly with increase of the number of qubits and cluster state dimension. Qumodes overcome these issues and for a CV optical system, qumodes, offers rich degrees of freedom and are able to deterministically generate the entanglement. 

In the rest of this survey, we discuss in details the various approaches adopted in physically realizing MBQC (in both CV and DV) and report on their experimental results.

\section{CV-Based Quantum Computation}\label{CVQC}
Quantum computation based on continuous variables relies on the promising attribute of optical parametric oscillators (OPOs) which are capable of producing  significantly large number of quantum fields \cite{pfister:2004, Menicucci:2008}, ranging from thousands to millions of quantum modes, usually termed as qumodes \cite{Menicucci:2006,Weedbrook:2012, Furusawa:2011} in time \cite{Yokoyama:2013,Yoshikawa:2016}, and frequency \cite{Pysher:2011,Chen:2014,Wang:2015, WangPei:2014} domain multiplexing. CV-based quantum computation was proposed in 1999 by Lloyd and Braunestein \cite{Lloyd:1999}, which improves fault tolerance and quantum error correction \cite{Menicucci:2006,Menicucci:2014}, while maintaining the computational advantage of universal quantum computation.

\paragraph{State representations}
Conventional quantum computation uses Discrete Variables (DV), where qubit is the basic information unit. The qubit is a two-level system, i.e., a two-dimensional Hilbert space with computational basis states $\ket{0}$ and $\ket{1}$. The conjugate bases states for computational basis states are $\ket{+}$ and $\ket{-}$. These two bases are related by Hadamard operation H.
In CV based quantum computation the basic information unit is qumode\footnote{We will also often refer to qumode as entangled modes, modes or flying qubits in this paper} \cite{Mile:2009}. Unlike qubit, qumode has Hilbert space with infinite dimensions, which is spanned by continuum of orthogonal states $\ket{s}_q$, where each $s \in \mathbb{R}$. The orthogonality condition is $\bra{r}_q\ket{s}_q=\delta(r-s)$. The conjugate basis states are labelled as $\ket{s}_q$.
The relation between the two bases $\ket{s}_q$ is represented by Fourier transform operation, in (\ref{conj_bases_eq1}). 
\begin{equation}
    \begin{aligned} \label{conj_bases_eq1}
    \ket{s}_p=\frac{1}{2\pi} \int_{-\infty}^{\infty} dre^{-irs} \ket{r}_q = F\ket{s}_q    \\ \ket{s}_q = \frac{1}{2\pi} \int_{-\infty}^{\infty} dre^{-irs} \ket{r}_p = F^{\dagger} \ket{s}_p
    \end{aligned}
\end{equation}
Equation (\ref{conj_bases_eq1}) also defines the unitary operator $F$. Qubits can be directly encoded as qumodes to perform quantum computation, such as coherent state encoding \cite{Ralph:2003}, or GKP encoding \cite{Gottesman:2001}. However, qumodes can also be used directly for any kind of CV based quantum computation \cite{Lloyd:1999}. The corresponding observables in CV-based quantum computation can be defined as momentum $\hat{p}$ and position $\hat{q}$, where $\hat{p}\ket{s}_p = s\ket{s}_p$ and $\hat{q}\ket{s}_q = s\ket{s}_q$ with $[\hat{q},\hat{p}=\textit{i}]$, where -$\hat{q}$ generates the positive translations in momentum and positive translations in position are generated by $\hat{p}$. Hence, an arbitrary position and momentum eigenstate can be written as in (\ref{eq:2}), where $X(s)=e^{-\textit{i}s\hat{p}}$ represents displacements in computational bases and $Z(s)=e^{-\textit{i}s\hat{q}}$ represents displacements in conjugate bases. The superposition of either $\ket{s}_p$ or $\ket{s}_q$ can form an arbitrary quantum state $\ket{\phi}$ of a CV system. 

\begin{equation} \label{eq:2}
    \ket{s}_q = X(s)\ket{0}_q \quad,\quad \ket{s}_p = Z(s)\ket{0}_p 
    \end{equation}

The computational basis or its conjugate is not countable, however, an arbitrary physical state $\ket{\phi}$ can be decomposed into countable infinite basis. For instance, Fock basis of definite particle number $\{ \ket{0},\ket{1},....\}$can be used for quantum optical fields or particles in harmonic trap, where $\hat{n}=\hat{a}^{\dagger}\hat{a}$ is the number operator with $\hat{n}\ket{n}=n\ket{n}$, the usual bosonic operator $[\hat{a},\hat{a}^{\dagger}]=1$ and $\hat{a}=(\hat{q}+\textit{i}\hat{p})/\sqrt{2}$.

\paragraph{Vacuum and squeezed states}
In the context of quantum optics (which is the most widely used paradigm for the realization of CV based quantum computation), for a given mode, $\hat{p}$ and $\hat{q}$ represent the \emph{momentum quadrature} and \emph{position quadrature}, respectively \cite{Mile:2009}. By minimizing the quadrature deviations product ($\Delta p\Delta q=1/2$), the qumode state has the minimum uncertainty. The vacuum or ground state $\ket{0}$, defined by $\hat{a}\ket{0}=0$ is of significant importance here both theoretically and practically. it also represents the Gaussian superposition centered about $0$ in computational or conjugate basis.
\begin{equation}
    \ket{0}=\frac{1}{\pi^{1/4}} \int dse^{{-s^2}/2}\ket{s}_q = \frac{1}{\pi^{1/4}} \int dse^{{-s^2}/2}\ket{s}_p
\end{equation}
The quadratures of vacuum state exhibit gaussian statistics and hence is a typical example of Gaussian state. 
The Einstein–Podolsky–Rosen (EPR) states are CV equivalent of bell states in qubit-based cluster states \cite{Pfister:2019}, where the summation of bell states is transformed to indefinite integral, which provides an intuition that EPR states are unphysical because they have infinite energy. However, it is possible to experimentally realize an approximation of these states, usually referred to as two-mode squeezed (TMS) states \cite{Ou:1992}.
The OPAs can directly create such states, for instance a doubly resonant OPO below threshold \cite{Ou:1992,walls:2007}. However, for vacuum states, finitely squeezed variance can be achieved \cite{Pfister:2019}, and are used widely for cluster states realization in CV-based quantum computation.

An arbitrary Gaussian state is completely determined by the first and second moments of quadrature, and  Wigner function is a quite handy tool for its description. By definition, any state with Gaussian Wigner function is a Gaussian state. 
Wigner function \cite{Mile:2009} can efficiently describe the qumode states (however, since a cluster state realized by continuous variables is multi-mode state, a multi-mode Wigner function is used).

\paragraph{Cluster States in CV}\label{CV_clusterstates}
1WQC is based on cluster states which includes the entire entanglement resource needed for the computation, which is then governed by single qubit measurements. The canonical way of cluster states realization (using controlled phase (CZ) gate for entangling two neighboring qubits), in qubit-based quantum computation, are explained in Section \ref{cluster_prep} 
%
%

1WQC using CV based cluster states has been formulated in \cite{Menicucci:2006,Mile:2009}. Moreover, in a recent survey on CV based cluster states \cite{Pfister:2019}, a comparison between qumode and qubit-based quantum computation is presented, following which, the cluster state for 1WQC can be created quite easily for CV based quantum computation \cite{Zhang:2006} . The most common approach to create cluster states in CV is the application of CZ gate in phase quadrature eigenstates $\ket{p}=0$, along the qumodes square lattice \cite{Pfister:2019}. In CV analog of qubit cluster states $\ket{+}$ becomes $\ket{0}_p$. The $Z$ measurements are replaced by $\hat{q}$, and $X$ measurements are replaced by $\hat{p}$. The CV analog of qubit controlled phase gate is $CZ=e^{\textit{i}\hat{q_i}\hat{q_j}}$, used for entangling the nodes $i$ and $j$. 

Such a cluster state in CV cannot be realized physically since they are infinitely squeezed, and practically only finitely phase-squeezed states can be physically implemented. To create these states, the degenerate optical parametric amplifiers (OPA), also known as single-mode squeezers are used, where quantum non-demolition operation is applied \cite{Caves:1980,yurke:1985,Miwa:2009}, which can also be called as controlled phase operation. 

\paragraph{Stabilizer Formalism}
\label{nullifier_section}

Stabilizer formalism for CV systems, also known as nullifiers, \cite{Barnes:2004,Loock:2007} can completely specify the CV based graph states \cite{Zhang:2008}, just like in the case of qubit cluster states. For instance, $X(s)$ is +1-eigenstate of momentum operator, it can stabilize the zero-momentum state $\ket{0}_p$. The stabilizers can be generalized for any CV based graph state $\ket{\varphi}$, with $n$ modes and graph represented as $G = (V,E)$.
Furthermore, EPR entangled states are the joint eigen states of two-mode commuting quantum variables \cite{Bohr:1935} used in quantum optics as described in (\ref{amplitute_quadrature_eq}) and (\ref{phase_quadrature_eq}) ($P1+P2$ and $Q1-Q2$). The plus and minus signs can be interchanged, which means that quantum standard deviation and noise measurement for these operators will be zero ($\Delta(Q1-Q2)=0$ and $\Delta(P1+P2)=0$). These EPR operators are also known as nullifiers or variance-based entanglement witness, in CV-based quantum computation \cite{Pfister:2019}, and is used widely for entanglement verification in CV based cluster states.

\begin{equation} \label{amplitute_quadrature_eq}
    P = \frac{1}{\textit{i}\sqrt{2}}(a-a^\dagger)
\end{equation}

\begin{equation} \label{phase_quadrature_eq}
    Q = \frac{1}{\textit{i}\sqrt{2}}(a+a^\dagger)
\end{equation}

\paragraph{Optical Implementation}
Quantum optics is the most natural platform for the implementation of CV-based quantum computation using EPR entanglement \cite{Einstein:1935}, which is believed to be the most versatile entanglement resource for various QIP protocols \cite{nielsen:2002,Furusawa:2011}, with various applications, including quantum dense coding \cite{Li:2002}, quantum key distribution \cite{Grosshans:2003}, unconditional quantum teleportation \cite{Fursawa:1998} and quantum secret sharing \cite{lau:2013}. Moreover, implementing CNOT (a universal quantum gate) was demonstrated on quantum optics \cite{Yoshikawa:2008}. 
In quantum optical implementations of CV-based quantum computation, the amplitude and phase quadrature operators of quantum electromagnetic field, as shown in (\ref{amplitute_quadrature_eq}) and (\ref{phase_quadrature_eq}), are the primary quantum variables for the underlying quantum computation \cite{Pfister:2019}, and are the mathematical equivalent of momentum and position operators of quantum harmonic oscillator of annihilation operator. The term \emph{quadrature} encapsulates the free evolution of harmonic oscillator's momentum and position.
The corresponding quantum gates and operations in CV-based quantum computation to that of qubit-based quantum computation are presented in a recent survey \cite{Pfister:2019}.


\paragraph{Inseparability criteria in CV Cluster states.}
The state nullifiers as discussed above in this section are linear combinations of momentum and position operartors for which the cluster states are eigenstates with 0 eigenvalue. The uncertainty measurement of the state nullifiers typically determines the inseparability in multi-partite cluster states \cite{Larsen:2019}. However, when it comes to actual experimental realization, the cluster states do not behave as exact eigenstates of nullifiers, and measurement results in uncertainties around zero.  According to van Loock-Furusawa criteria of inseparability \cite{Loock:2003}, two or more sets of modes for a given cluster state are formed and an inequality is then derived with combined quadrature variance (usually should be $<$ -3dB for linear CV cluster states). Any violation in that inequality results in failure of inseparability criteria for the underlying cluster state, and might result in failure of entanglement detection – an imperative feature of cluster states. Hence, the initial degree of quadrature squeezing directly effects the entanglement of quantum oscillator being exploited to make cluster state. Among others, a relatively recent work \cite{Korolev:2018}, highlights the minimum squeezing criteria for cluster state generation.

\paragraph{Scalability of CV cluster states:} \label{scalability in_CV_cluster_states}
Scalability is at the heart of universal quantum computation. From the scalability viewpoint of canonical method of cluster states generation in CV domain, $N$ number of degenerate OPAs are needed for an $N$-mode cluster state with approximately two online OPAs (squeezers), for a single entangling operation \cite{Braunstein:2005}. This approach is often termed  “bottom up” approach in the literature.  The experimental resources increase linearly with the size of cluster state to be created and hence not very scalable \cite{Pfister:2019}. After the realization of the fact that any N-mode graph state with arbitrary accuracy $s$, is a multi-mode Gaussian resource \cite{Loock:2007,Braunstein:2005}, the canonical cluster state creation protocol can be re-cast as a Bloch–Messiah Reduction (BMR) \cite{Braunstein:2005}. The BMR approach allows to decompose a complex Gaussian unitary into simple scheme, where linear optical components can be separated from non-linear components. The cluster state creation approach, by exploiting the BMR technique, is usually termed as decomposition or top-down approach. The decomposition approach removes the requirement of two online squeezers for entangling operation in canonical approach of cluster state creation by sandwiching an $N$-single mode OPAs between the two N-mode interferometers\cite{zhu:2020}. The cluster state can then be created with offline squeezing (vacuum state squeezing) and linear optics, which in fact is more advantageous since the online squeezing (squeezing of an arbitrary state of electromagnetic field ) is experimentally more challenging than offline. The first interferometer becomes irrelevant in case of vacuum states as input states and a single N-mode interferometer can effectively realize such Gaussian states, which significantly simplifies the cluster state creation protocol by using the linear optical interferometer instead of optical controlled phase gates \cite{Loock:2007,Yukawa:2008}. Furthermore, it has been proved in \cite{Larsen:2019}, that offline squeezing saves significant amount of dBs as compared to online squeezing and hence increasing the number of qumodes being generated.

As discussed, the entangling operation’s requirement of two OPAs has already been solved via the top-down approach, however the use of $N$ number of OPOs for scaling up to N-mode cluster state is still a major hurdle in the context of experimental realization of CV-based cluster states, constraining the scalability of CV cluster states. Another improvement on scaling CV-based cluster states is the idea of using the whole quantum optical frequency comb (QOFC) of a single OPO for scaling up to $N$-mode CV cluster state rather than using $N$ number of OPO’s for scaling up to $N$-mode CV cluster state \cite{Pfister:2019}. Moreover, the idea of using whole QOFC of single OPO also surpassed the requirement of interferometers, and hence providing a great potential for scaling the cluster states \cite{Menicucci:2008}. The first implementation exploiting the N-mode squeezing for the generation of N-mode entanglement was only for the GHZ state \cite{pfister:2004}, which was then extended to cluster state generation \cite{Menicucci:2007}, with square lattice configuration using a single OPO \cite{Menicucci:2008}.



\subsection{Experimental Realization of CV Cluster states} \label{CV_exp_realization}
Quantum optics provides a scalable platform for universal CV-based quantum computation. This has been demonstrated with frequency and temporal phase-locked quantum optical combs, which are emitted by OPOs \cite{Menicucci:2006,Menicucci:2008,Pfister:2019,Flammia:2009}. Moreover, it has been shown in \cite{Zhang:2006}, that cluster states are generated upon interference of shifted, two-mode-squeezed optical combs. These states are the universal resources for quantum computation \cite{Raussendorf:2001}, in time and frequency domains, discussed later in this section.  The two-mode-squeezing states, sometimes known as EPR pairs, are the building blocks of a CV cluster state \cite{Pfister:2019}. Entangling qumodes from different EPR pairs results in cluster state chain, also called quantum wire. Recently, the optical spatial comb has also been used to realize CV cluster states using spatial multiplexing, also discussed later in this section. In the following subsections, we will discuss state-of-the art CV cluster states using temporal, frequency and spatial domains. Furthermore, hybrid approaches, using a combination of any two of the multiplexing approaches, are also discussed.


\subsubsection{Using Frequency Domain Multiplexing}\label{CV_FDM}

\begin{figure}[htp]
	\centering
     \begin{subfigure}[b]{0.4\textwidth}
         \centering
         \includegraphics[scale=0.35]{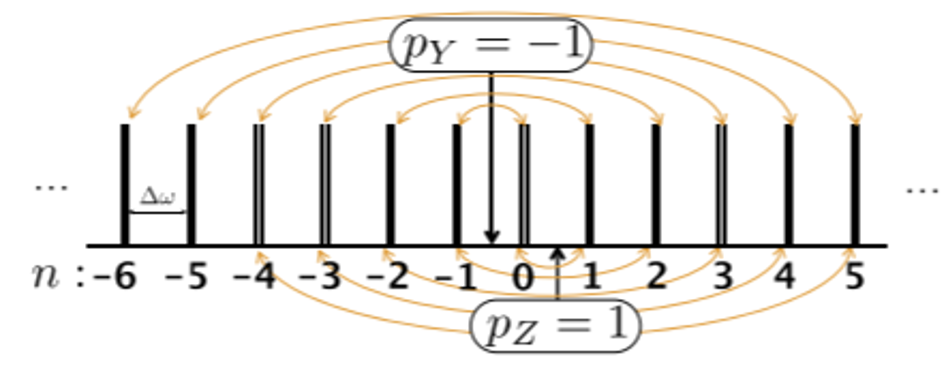}
         \caption{}
         \label{fig:F1_a}
     \end{subfigure}
    \hfil
     \begin{subfigure}[b]{0.4\textwidth}
         \centering
         \includegraphics[scale=0.3]{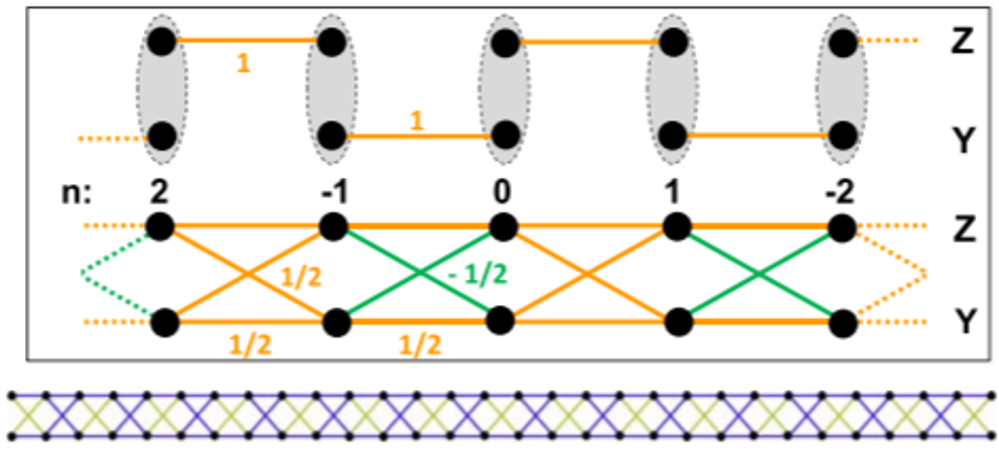}
         \caption{}
         \label{fig:F1_b}
     \end{subfigure}
	\caption{Generation of CV Dual-rail Quantum wire in Frequency Domain \cite{Chen:2014}: (a) Initial graph in quantum optical frequency comb. The arrow indicates pumps’ half frequencies. (b) A chain structure is formed by the recorded frequencies. The ovals represent the balanced beam splitter interactions. The measured 60 qumode CV cluster state is at the right panel of (b).}
	\label{fig:F1}
\end{figure}

In frequency domain realization of CV cluster states, the entangled modes are simultaneously generated \cite{Chen:2014}. 
In frequency domain the input EPR pairs straddle many frequencies and two orthogonal polarizations, created in two sets at two orthogonal linear polarizations. The pairs at one polarization are shifted with respect to the pair at another polarization by frequency shifting of the independent pump fields that creates each pair set, Fig. \ref{fig:F1_a}. This frequency shifting is a lossless operation. 
All EPR pairs are emitted in the same cavity mode and are subjected to balanced beam splitting by undergoing a $45^\circ$ polarization rotation in a halfwave plate resulting in a dual-rail quantum wire, as shown in Fig. \ref{fig:F1_b}.

A prominent advantage of frequency domain is the lossless implementation of large delays, which are needed for scaling up to large number of wires, as compared to implementation in time domain where optical fiber delay lines are usually used. However, the frequency domain implementation is limited by the phasematching bandwidth whereas time domain implementation is limited by characteristic stability time of experiment under consideration, which has no fundamental restriction \cite{Pfister:2019}. 

A limitation of frequency domain is the possibility of running out of the phasematching bandwidth due to OPO crystals dispersion. In such a case the QOFC’s FSR’s become chirped and qumodes which are far away from the pump’s half-frequency will shift out of OPO resonance \cite{Pfister:2019}. A proper use of spectrally broadened pump field can help overcoming this very problem \cite{Wang:2015}.


Some initial proposals for the generation of cluster states based on continuous variables using frequency domain multiplexing \cite{Menicucci:2006,Yokoyama:2013}, involved the use of online squeezers (seeded OPOs \cite{Menicucci:2008}), which are quite challenging to implement. However, as discussed earlier in section \ref{CV_clusterstates}, a more practical approach is the use of Bloch-Messiah decomposition, which makes use of $N$ number of vacuum squeezers along with $\mathcal{O}(N^2)$ port interferrometers \cite{Loock:2007}. This very method also uses significant number of experimental resources and hence offer limited scalability. To overcome this issue in terms of scalability of CV-based cluster states, a new approach for the creation of larger cluster states, having a compact experimental setup is proposed in \cite{Menicucci:2008}, and is also called \emph{top-down} approach. This approach offers a promising scaling potential since it only uses a single OPO and no interferrometer. 

\begin{figure}[!htp]
	\centering
		\includegraphics[scale=0.5]{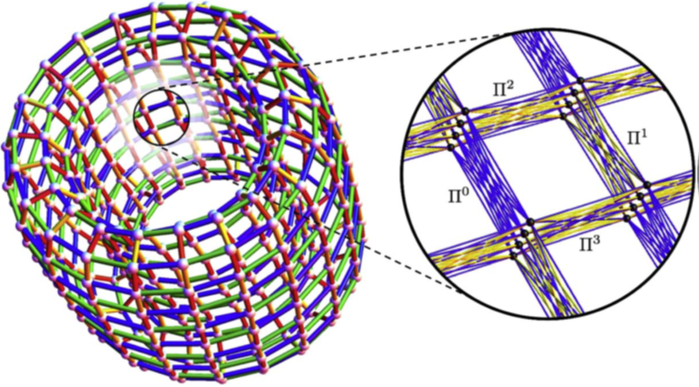}
	\caption{Toroidal cluster state generation in Square lattice configuration \cite{Menicucci:2008}. (Left): The resulting graph for CV cluster state. (Right): each white vertex comprises of a set of four individual qumodes which are labelled by two orthogonal polarization and two frequencies.  The non-linear interactions (ZYY,ZZZ and YZY/YYZ) are represented by blue and yellow edges respectively. } 
	\label{fig:F3}
\end{figure}

The initial proposal on square-lattice cluster state creation exploiting a single OPO was presented in \cite{Menicucci:2008,Flammia:2009} and is shown in Fig. \ref{fig:F3}, which overcame the no-go theorem in linear chain and square lattice cluster states creation using QOFC \cite{Flammia:2009}. The proposal expanded the proof (cluster states creation using single OPO by using its whole QOFC), by the addition of an extra degree of freedom (polarization) to the qumode’s frequency labels. A doubly resonant OPO containing a periodically poled KTiOPO4 (KTP) crystal with phase matching the three different pump/signal/signal polarization sets ZZZ, ZYY and YZY/YYZ, all with same coupling strengths was used for the implementation of polarization block. The experimental demonstration of this crystal is presented in \cite{Pysher:2010}. 

The non-trivial frequency spacing and a fairly complicated 15-mode pump field with orthogonal $\pm 45^\circ$ polarization components is slightly inconvenient in this proposal, which needs sophisticated phase modulation techniques for producing the single side bands \cite{Izutsu:1981} at multiple frequencies. This inconvenience led to the exploration of other techniques for cluster state generation, it is still, however compact and might be implementable in future. 


An experimental implementation of proposal \cite{zaidi:2008}, for multiple $2\times2$ cluster state creation is presented in \cite{Pysher:2011}. Despite having  smaller cluster size, it exhibits a great scalability potential because of the simultaneous generation of 15 copies of square lattices with each having 4 qumodes and hence a total of 60 qumodes cluster state. However, the limited number of qumodes in the resulting state are primarily due to the experimental limitations and more sophisticated setup would lead to at least three times larger cluster state \cite{Pysher:2011}. This was the first large scale cluster state generation. The OPO has two KTP crystals where one was phasematched the ZZZ and YZY/YZ interactions simultaneously and is coupled with two frequencies and two polarizations with a single pump frequency. 


While using the QOFC, the scalability of a cluster state is not only dependent on the number of qumodes (state size) but also on the number of copies of the state \cite{Pfister:2019}. To see this in frequency domain (Fig. \ref{fig:F1_a}), we detune the pump half-frequencies by an integer multiple of the OPO FSR . This idea has been demonstrated in \cite{Chen:2014}, where, by using a single OPO result in two independent quantum wires each having 30 qumodes. The same proposal also implements a single dial-rail quantum wire with simultaneously accessible 60 qumodes (Fig. \ref{fig:F1}).


An $N\times N$ square lattice in frequency domain is proposed in \cite{WangPei:2014}, where two QOFC’s were interfered. One QOFC was hosting half pump detuning of 1 FSR (single wire whereas the other QOFC was hosting the half pump detuning of $N$ FSRs ($N$ independent wires). This proposal was based on the original temporal idea of \cite{Menicucci:2011}, but it opened doors to new type of scalability where, by using only one QOFC per dimension and generalized interferometers in a fractal procedure \cite{WangPei:2014}, we can go from 1D to 2D and then 2D to 4D.

Most of the above-mentioned proposals depend on the interference of two to four squeezed quantum frequency combs. Recently, a frequency domain realization of CV cluster states with more compact experimental configuration was proposed \cite{zhu:2020}, where only a single comb is shown to be sufficient for the generation of n-hypercubic cluster states having arbitrary dimension $n$. Moreover, cluster states with 1D, 2D and 3D have also been realized. In \cite{zhu:2020}, phase modulation via sparse discrete spectrum is performed on QOFC generated by a single OPO. The size of cluster state is dependent on the spacing between the modulation frequencies and the dimension of cluster state is determined by the total number of modulation frequencies. First, independent EPR qumodes pairs in TMS states, also known as EPR pairs, are generated over QOFC by a single doubly resonant OPO with a single pump frequency evenly distributed between two OPO mode frequencies following the procedure as in Fig. \ref{fig:F1_a}. Second, by optical or RF means, phase modulation is performed at frequency multiple of comb spacing.



\begin{figure}[!htp]
	\centering
		\begin{subfigure}[b]{0.4\textwidth}
         \centering
         \includegraphics[scale=0.4]{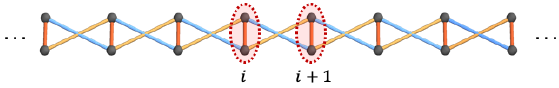}
         \caption{}
         \label{fig:F8_a}
     \end{subfigure}
     \qquad
     \hspace{2cm}
     \begin{subfigure}[b]{0.4\textwidth}
         \centering
         \includegraphics[scale=0.4]{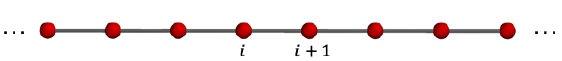}
         \caption{}
         \label{fig:F8_b}
     \end{subfigure}
	\caption{Generated 1D cluster state \cite{zhu:2020}. (a) the red ovals represents EPR\hyp{}
	qumodes containg two modes. (b) Compact Graph representation of 1D cluster states using EPR macronodes. }
	\label{fig:F8}
\end{figure}

The entanglement is scalable in terms of number copies of bipartite EPR states, but not in terms of size of multipartite state. It was shown in \cite{zhu:2020} that performing phase modulation of the OPO's QOFC by an EOM in 1, 2 and 3 frequencies will lead to the creation of 1D, 2D and 3D cluster states. For 1D cluster states, single phase modulation with 0, 0.1, 0.2, 0.5 and 1 radian, with an initial squeezing parameter of 2.3 was performed resulting in different configuration of 1D dual-rail cluster states. 
A typical structure of generated 1D graph is shown in Fig. \ref{fig:F8}. 

\begin{figure}[!htp]
	\centering
		\begin{subfigure}[b]{0.4\textwidth}
         \centering
        \includegraphics[scale=0.7]{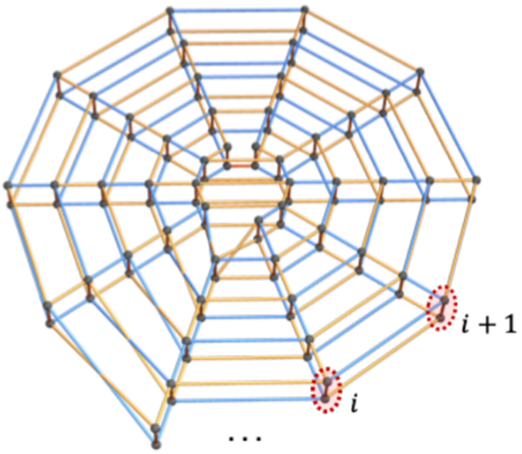}
         \caption{}
         \label{fig:F9_a}
     \end{subfigure}
     \hfil
     \begin{subfigure}[b]{0.45\textwidth}
         \centering
         \includegraphics[scale=0.3]{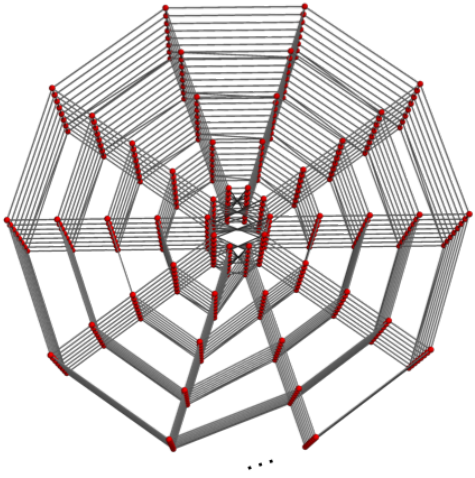}
         \caption{}
         \label{fig:F9_c}
         
     \end{subfigure}
   
	\caption{Generated 2D and 3D cluster states \cite{zhu:2020}. (a) the red ovals represent the EPR qumodes, each containing two modes with one mode in top layer and one in the bottom layer. (b) The generated 3D cluster state in cylindrical configuration} 
	\label{fig:F9}
\end{figure}

Phase modulation at an additional frequency in the setup from Fig. \ref{fig:F8} yielded a 2D cluster state. The two phase modulation frequencies for 2D cluster states were $\Omega_1=1$, which created a ladder graph just like 1D case, and $\Omega_2=10$, which introduced additional coupling after every 10 modes, transforming the 1D graph to 2D, as shown in Fig. \ref{fig:F9_a}.

The width of square lattice (number of spokes) in the 2D case is determined by $\Omega_2/\Omega_1$. The total number of qumodes generated is determined by phasematching bandwidth of the non-linear medium in OPO. Addition of another modulation frequency lead to 3D cluster state generation, Fig. \ref{fig:F9_c}. The modulation frequencies for 3D cluster state were, $\Omega_1 = 1, \Omega_2 = 8$ and $\Omega_3 = 80 $. The length of cylindrical cluster state is set by $\Omega_2/\Omega_1$ and number of spoke in 3D case are set by $\Omega_3/\Omega_2$. An increase in the number of modes increases the spoke radius as $N/\Omega_3$. The example of generated 3D cluster state as shown in Fig. \ref{fig:F9_c}, has 400 macronodes with 10 spokes, 5 set of macronodes in radial direction and a length of 8 macronodes. All state-of-the-art approaches exploiting frequency domain multiplexing for cluster states realization are summarized in Table \ref{tab:summary_CV-Domain}.


\subsubsection{Using Time Domain Multiplexing:} \label{CV_TDM}
Another approach to realize CV cluster state is time domain multiplexing, which has seen considerable attention in the recent years. In time domain multiplexing, two-single mode squeezed states are interferred in quadrature to create spatially separated EPR pairs \cite{Pfister:2019}. A dual rail quantum wire structure is created by passing one qumode of one of the EPR pair through delay line before its interference with a qumode of next EPR pair at a beam splitter.  The experimental realization of this idea is presented in \cite{Yokoyama:2013}, with $10^4$ qumodes in a single chain. This idea was later extended to generate around one million qumodes that are sequentially accessible (two qumodes can be accessed simultaneously \cite{Yoshikawa:2016}). It is important to note that even the sequential access of qumodes is compatible with quantum computation and is termed as the \emph{Wallace and Gromit approach}  \cite{Menicucci:2010}. 




The top-down and bottom-up approaches are also being used for cluster states realization in time domain. When using the whole QOFC of a single OPO while scaling the CV cluster state, the scalability is not only dependent on the number of modes per state (state size), but also on the number of copies of the state. 
For the realization of square or hypercubic cluster states in temporal domain, two commensurate delays are used to come up with a square lattice \cite{Pfister:2019}. Although, such temporal delays pose a major limitation for the scalability of cluster states, the experimental implementation of cluster states using this approach has progressed well recently \cite{Larsen:2019,Asavanant:2019}. 

A deterministic generation of continuous variable cluster state, called extended Einstein-Podolsky-Rosen (XEPR) state, along with its full characterization is presented in \cite{Yokoyama:2013}, which enables the ultra-large scale QIP based on MBQC, since the generated cluster state contains more than 10,000 entangled modes. These entangled modes are time domain multiplexed wave-packets of light. Combining two XEPR states with different time delays using additional beam splitters can help achieve universal MBQC \cite{Mile:2009}. The primary reason behind the creation of large number of entangled modes is that only a small subset of the whole XERP state exists at each instant of time in the time domain multiplexed demonstration. The sequentially propagating EPR states which are contained in two distinct beams were entangled for the generation of XEPR states. 

        

Soon after the realization of XEPR cluster state containing $10,000$ entangled modes, a CV-based cluster state with around one-million modes was developed \cite{Yoshikawa:2016}. The authors highlighted and addressed the issues in XEPR cluster state realization in \cite{Yokoyama:2013}, such as the the long optical delay line in the optical setup, and the noisy modulated bright beams used for phase locking (necessary for cluster state generation). These modulated beams treat interference signals as error signals for feedback control. All these issues lead to failure in inseparability criteria \cite{Loock:2003}, after 16000 entangled modes in 1.3ms \cite{Yokoyama:2013}. To overcome these issues, continuing feedback control strategy for the optical system was used in \cite{Yoshikawa:2016} during cluster state generation and electrically removing the noise of modulated bright beams helped in achieving more than one million entangled modes without any degradation in squeezing criteria. The inseparability criteria for enough entanglement witness between the qumodes is calculated to be -3 dB. However, the qumodes up to $6\times10^5$ modes had nullifier variances well below the inseparability with  ${\hat{n}}_k^p=-4.3\pm 0.2 dB$ and ${\hat{n}}_k^x=-4.3\pm0.2 dB$. The worst variances reported for qumodes over $6\times10^5$ are ${\hat{n}}_k^p= -3.6 dB$ and ${\hat{n}}_k^x=-3.5 dB$. 



\begin{figure*}[!htp]
	\centering
		\includegraphics[scale=0.35]{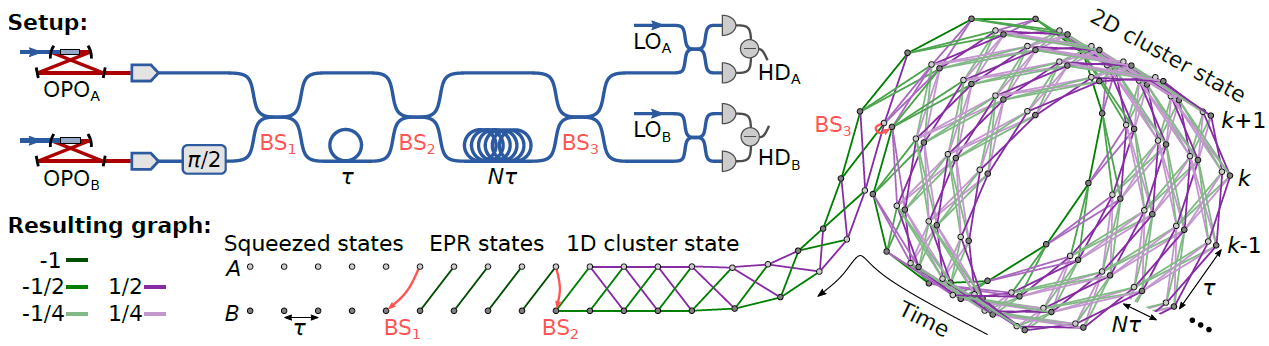}
	\caption{Generation Process of 2D cluster state \cite{Larsen:2019}. The $OPO_A$ and $OPO_B$ produce squeezing. The EPR states are created by the interference of the temporal modes of squeezing with mode index $k$ in two spatial modes A and B at beamsplitter $(BS_1)$. The mode B is then delayed by $\tau$ and EPR pairs get entangled at beamsplitter $(BS_2)$ to form a 1D cluster state. The mode B is again delayed by $N_\tau$ and the 1D cluster state is curled up to form a 2D cluster state at beamsplitter $(BS_3)$. The homodyne detectors $HD_A$ and $HD_B$ are used for measuring the temporal mode quadratures which then aid towards nullifiers calculation.} 
	\label{fig:T4}
\end{figure*}

The cluster states realizations discussed above are 1D and/or dual rail. However, for universal quantum computation, multi-qubit (or qumodes in CV case), operations are imperative, which usually needs cluster states having two or more dimensions \cite{Menicucci:2011,Economou:2010,WangPei:2014}. A two-dimensional cluster state is usually a square lattice configuration of qumodes, which has been realized in CV using time domain multiplexing \cite{Asavanant:2019} \cite{Larsen:2019}. 
%
The position and momentum quadrature’s of photonic harmonic oscillator \cite{Weedbrook:2012}, were used for information encoding. A long chain of entangled modes were generated in \cite{Larsen:2019} by temporal multiplexing of optical Einstein-Podolsky-Rosen (EPR) states \cite{Reid:2009}, which was then curled and fused into a two-dimensional (2D) cylindrical array of qumodes. An abstract experimental setup demonstration along with the generated cluster state is shown in Fig. \ref{fig:T4}.

A total of 30,000 entangled modes are generated in \cite{Larsen:2019} with $2\times12=24$ modes as input registers (each having 1250 modes), which can be used for encoding the input state. The inseparability bound for entanglement creation between qumodes was found to be -3dB and is satisfied by the qumodes of generated cluster states, as presented in tabular summary at the end of this section. Furthermore, the proposed approach in \cite{Larsen:2019} is quite flexible for upscaling mainly because of its deterministic generation and easy-to-achieve experimental conditions in optical fibers.



Another recent experimental implementation of large scale 2D CV cluster state is presented in \cite{Asavanant:2019}, which is capable to implement universal MBQC in CV domain.
The total number of qumodes generated are around 25000 with 5 input modes and a computation depth of 5000 modes. Multiple temporally localized square shaped cluster state on four beams which are spatially separated are used for generation of 2D CV time domain multiplexed cluster state by exploiting time domain multiplexing. Two optical delay lines are used for creating the 2D structure. The delay of on optical delay line is same as time interval between temporal modes of the state. The delay of other line is equal to the product of number of input modes for quantum computation and time interval between temporal modes. Finally, a 2D cluster state is created by first delaying the modes on two beams and then connecting them to temporal modes on two non-delayed beams. An example cluster state for 30 input modes is similar to that created for frequency domain multiplexing (Fig. \ref{fig:F3}). The inseparability criteria for the generated qumodes is found to be less than -4.5 dB and is fulfilled.



More recently, a novel approach for the generation of large-scale three-dimensional (3D) topological cluster state, was proposed in \cite{Fukui:2020}, providing a platform for topologically protected MBQC. The proposed approach combines the time domain multiplexing and divide-and-conquer approach \cite{nielsen:2004,Dawson:2006}, resulting in two prominent advantages. Firstly, for entanglement verification of the created cluster state, the squeezing levels are quite feasible in terms of experimental implementation, because the squeezing level needed for the proposed 3D cluster state is $\simeq$4.77dB, which does not exceed much from that of 2D cluster state exploiting only the time domain multiplexing, which is recently reported in \cite{Asavanant:2019} and is $\simeq$4.5dB. Secondly, the generated cluster state exhibits promising robustness against analog errors. These two advantages make it an appropriate choice for large scale quantum computation in CV domain.   
All the state-of-the-art techniques using time domain multiplexing for cluster states generation are summarized in Table \ref{tab:TDM_table}.





\begin{table}[!htp]
    \centering
    \caption{State-of-the-art cluster state realization using time domain multiplexing}
   \begin{tabular}{p{3.8cm} p{0.8cm}| p{2.0cm}| p{2.2cm}| p{2.3cm}| p{2.6cm}}
\hline
Ref \# & & \hfil \cite{Yokoyama:2013} &\hfil \cite{Yoshikawa:2016} &\hfil \cite{Larsen:2019} &\hfil \cite{Asavanant:2019}  \\
\hline
Cluster state dimension &  & \hfil 1D-dual rail & \hfil 1D-dual rail & \hfil 2D  &\hfil 2D\\
\hline
Computation depth &  &\hfil- &\hfil - &\hfil 1250  &\hfil 5000\\
\hline
No.of qumodes &  &\hfil 10,000 &\hfil $1.2\times 10^6$ &\hfil 30,000 &\hfil $\simeq$25,000 \\
\hline
No.of squeezed light sources &  &\hfil 2 &\hfil 2 &\hfil 2 &\hfil 4 \\
\hline
No of beam splitters &  &\hfil 2 &\hfil 2 &\hfil 3 &\hfil 5 \\
\hline
No of optical delay lines &  &\hfil 1 &\hfil 1 &\hfil 2 &\hfil 2 \\
\hline
Qumodes inseparability bound &  &\hfil -3 dB &\hfil -3 dB &\hfil -3 dB &\hfil -4.5 dB \\
\hline
Multiplexing time &  &\hfil 157.5$ns$ &\hfil 160$ns$ &\hfil \makecell{371$\mu s^*$ \\ $371ms ^{**}$} &\hfil 200$ns$ \\
\hline

Nullifier variances  &\multicolumn{1}{|c|}{${\hat{n}}_k^p$} 
&\hfil $-5.2\pm0.2$ dB
&\hfil $-4.3\pm0.2$dB$^1$ &\hfil\makecell{-4.3dB$^*$ \\ -4.4 dB$ ^{**}$}  &\hfil \makecell{$-5.34 \pm 0.0$6 dB$^\dagger$ \\ $-4.3 \pm 0.2$dB$^{\dagger\dagger}$} \\
\cline{2-6}
&\multicolumn{1}{|c|}{${\hat{n}}_k^x$} 
& $-4.9 \pm 0.2$ dB
& $-4.3 \pm 0.2$ dB$^1$ &\makecell{-4.3 dB$^*$ \\ -4.4 dB$^{**}$}  & \makecell{$-5.34 \pm 0.06$ dB$^\dagger$ \\ $-4.3 \pm 0.2$ dB$^{\dagger\dagger}$} \\
        
\hline
Output sampling rate & &\hfil 200 MHz &\hfil 100 MHz &\hfil 250 MHz &\hfil 1 GHz \\
\bottomrule

\multicolumn{6}{l}{\footnotesize $^1 first 6\times 10^5 qumodes$}\\
\multicolumn{6}{l}{\footnotesize $^* Small Dataset (1500 qumodes)$}\\
\multicolumn{6}{l}{\footnotesize $^{**} Large Dataset (15000 qumodes)$}\\
\multicolumn{6}{l}{\footnotesize $^\dagger First delay line$ }\\
\multicolumn{6}{l}{\footnotesize $^\dagger\dagger Second delay line$}
\end{tabular}

\label{tab:TDM_table}
\end{table}


\subsubsection{Using Spatial Multiplexing}\label{CV_SDM}
Analogous to exploitation of optical frequency comb in frequency domain multiplexing approaches for cluster states realization, the spatial mode multiplexing exploits the spatial mode comb, which is slightly different yet an effective approach for large-scale cluster states realization. 
The spatial freedom of light is quite effective for up scaling the number of entangled modes (cluster state) \cite{Liu:2016} and can bring new improvements and extensions in cluster states physical realization. Spatial multiplexing offers some advantages which can lead to large scale generation of cluster states. First and foremost, in frequency multiplexing the small interval of frequency comb is dependent on free spectral range (FSR) of optical cavity, which makes it experimentally quite challenging to spatially separate them, whereas spatial modulators can quite efficiently spatially separate the modes in case of spatial modes \cite{Zhang:2017}. Moreover, as opposed to temporal and frequency modes where experimentally challenging local light fields with accurate frequency and more measurement times are prepared for modes detection, the detection of spatial modes is relatively simpler and can be detected by multi-quadrant detectors  \cite{Zhang:2017}. Spatial multimode entanglement has already been explored in \cite{Midgley:2010,Santos:2009} by exploiting a single multimode OPA with the generation of spatial quadripartite GHZ entanglement \cite{Liu:2016} and CV hyper entanglement in \cite{Santos:2009,Kui:2014}. 
%

As discussed, multiplexing in time or frequency domain are widely used for CV cluster states realization. The multiplexing is performed in OPOs both below and above threshold. A combination of entangling operators and beam splitter transformations governs the cluster states realization in these approaches. A similar transformation exists for OPAs with Gaussian input states that are operating on multiple spatial modes, where a careful selection of local oscillators leads to spatial mode distribution similar to that of optical frequency comb having axial modes in OPO cavity \cite{Pooser:2014}, which can then be exploited for cluster states generation. 

The first ever theoretical proposal to use multi-spatial mode amplifier configuration capable of generating dual-rail cluster states over optical spatial comb was proposed in 2014 in \cite{Pooser:2014}. That scheme uses insensitive amplifiers with concurrent phase based on four-wave mixing in alkali metals vapors. Every concurrent amplifier operates on independent spatial modes.  A careful selection of local oscillator for entangled spatial modes measurement then helped to generate a spatial frequency comb from amplified spatial modes. These spatial modes are then mixed via linear transformation for the generation of the cluster state. The primary focus in \cite{Pooser:2014} was to develop an analogy between optical frequency comb and optical spatial mode comb, which was then used for a dual rail cluster state realization. The proposed setup can be used to generate and detect cluster states using images to synthesize appropriate local oscillators. The proposed setup offers various advantages, such as ease of alignment, simple phase control and scalability using multiple gain regions. However, a potential disadvantage for that scheme is any local oscillator’s misalignment can introduce noise due to the use of multiple OPOs rather than a single OPO. 

In 2016, an experimental scheme for the generation of cluster states based on spatial mode combs by using a large-Fresnel-number degenerate optical parametric oscillator (DOPO) was proposed in \cite{yang:2016}. An eleven-partite dual-rail cluster state was generated, similar to what is depicted in Fig. \ref{fig:F1}. Two spatial Laguerre–Gaussian (LG) modes with the same frequency were used for DOPO pumping ($\lg_l^p$). The non-linear crystal of type I-phase matching ($\chi^2$) was used in the cavity assuring the sustainable and simultaneous non-linear interaction between all down-converted and pump modes. 


In 2017, the same group of researchers who proposed \cite{yang:2016}, came up with a new scheme for generating large-scale CV-based dual rail cluster state with more than 20 qumodes, by using a specially designed self-imaging OPO \cite{Zhang:2017}.  The proposed scheme with special OPO design is capable of multiplying the number of qumodes. It exploits the spatial mode comb in a self-imaging OPO. Two spatial Laguerre-Gaussian modes with same frequency and different polarization are used for pumping the OPO. 
The experimental setup uses two polarized type-zero phase matching nonlinear crystals are place within four-mirror ring cavity and EPR pairs are generated using PDC, which were then concatenated and in turn extends to optical spatial mode comb and are connected by curved arrows, similar to Fig. \ref{fig:F1_a}. This optical spatial mode comb is then passed from a single beam splitter resulting in a large scale dual-rail cluster state, in a similar way as the one illustrated in Fig. \ref{fig:F1_b}. The Van Loock Fursawa criteria of qumodes’ full inseparability \cite{Loock:2003}, is used for the evaluation of entanglement and it is shown that entanglement exists over a vast range of pump parameter and analyzing frequency. 


The entanglement between co-propagating modes in one beam using spatial domain has previously been demonstrated in \cite{janousek:2009}. The cluster state was also created where qumodes are defined as combinations of different spatial regions of one beam \cite{armstrong:2012}. More recently, in 2018, a CV square cluster state was developed by multiplexing the orthogonal spatial modes in single OPO \cite{Cai:2018}. Moreover, the pump profile is optimized by separately controlling the temperature of non-linear crystals inside the OPA cavity, during the experiment, which significantly improved the entanglement quality. Two first order Hermite-Gauss (HG) modes with one beam in a single multi-mode OPA is used for creating the multimode entanglement, which is then transformed into a cluster state by phase correction. This approach is scalable for multimode entanglement in spatial domain, and eventually to spatial cluster states, which are basic resource for spatial quantum information processing.  
The total efficiency of the experimentation process is determined by product of propagation efficiency ($n_{prop}=0.96\pm 0.02$ dB), photodiode efficiency ($n_{phot}=0.92\pm0.02$ dB) and the efficiency of spatial overlap in the homodyne detector ($n_{hd}=0.96\pm0.02$ dB).



All the state-of-the-art approaches exploiting spatial domain multiplexing for cluster states realization are summarized in Table \ref{tab:summary_CV-Domain}.


\subsubsection{Hybrid Approaches}

\begin{figure}[htp]
	\centering
		\begin{subfigure}[b]{0.3\textwidth}
         \centering
         \includegraphics[scale=0.3]{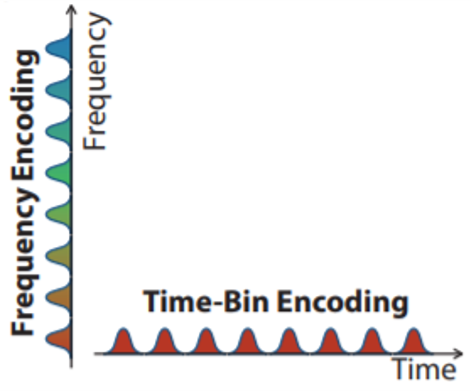}
         \caption{}
         \label{fig:H1_a}
     \end{subfigure}
     \quad
     \hspace{0.5cm}
     \begin{subfigure}[b]{0.3\textwidth}
         \centering
         \includegraphics[scale=0.3]{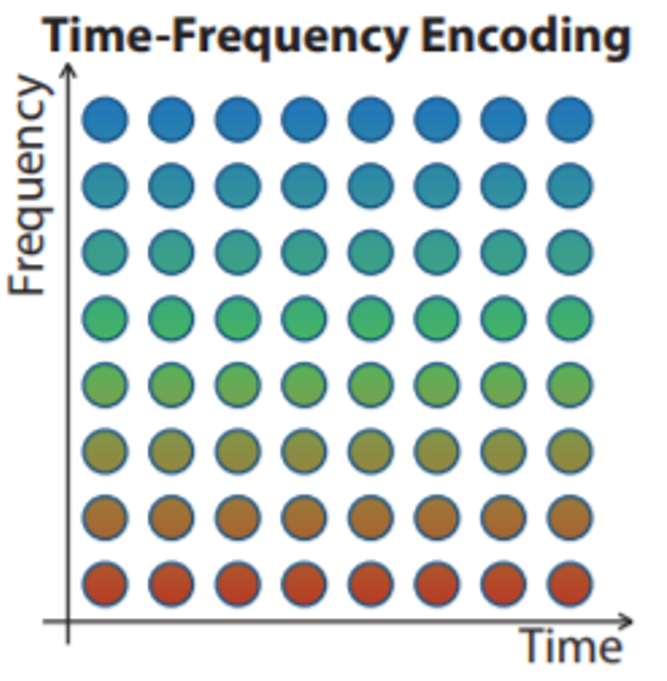}
         \caption{}
         \label{fig:H1_b}
         \end{subfigure}
	\caption{ Time, Frequency and Hybrid (Time and Frequency) encoding \cite{Peter:2014}: (a) Separate frequency modes and time-bin modes encoding of quantum information (b) Combined time and frequency encoding leading to a maximum channel capacity (qumodes) and eventually more quantum information} 
	\label{fig:H1}
\end{figure}

All the previous techniques, alone, can be visualized as 1D set of entangled modes; for instance, frequency and time domain multiplexing\cite{Peter:2014} as shown in Fig. \ref{fig:H1_a}. Although these approaches, single handedly can increase the number of qumodes and eventually the available information in an optical channel, a combination of these approaches can optimize channel capacity and hence provide more information by exploiting all the available encoding space \cite{Peter:2014}; Fig. \ref{fig:H1_b}.

Although, exploitation of these multiplexing techniques alone led to the creation of significantly large cluster states, a fully scalable experimental demonstration for 1WQC is still an open research problem. For that purpose, a combination of these approaches (temporal, frequency and spatial) is being used in order to obtain the best possible results. We now give an overview of some of the recent hybrid approached for CV cluster state realization.  

One of the earliest proposal using a combination of time and frequency domain multiplexing, based on quantum memories is presented in 2014 in \cite{Peter:2014}, for the cluster state generation. However, this approach follows the canonical method of CV cluster state generation (implementation of controlled-phase (CZ) gate for introducing the entanglement between generated qumodes) and is therefore not very scalable since CZ gate implementation in quantum optics is experimentally demanding.

Later in 2016, with independent developments in time and frequency domain CV cluster states realization, a hybrid approach combining frequency quantum wire generation from \cite{Chen:2014} and temporal entanglement from \cite{Menicucci:2011}, was proposed \cite{Alexander:2016}, where a cluster state is created by entangling the quantum frequency comb of EPR pairs by a single OPO, both in frequency and temporal domain. The resulting lattice could potentially contain infinite number of modes in one dimension (temporal modes), based on \cite{Yokoyama:2013} and up to $3\times10^3$ modes in the other dimension (frequency), based on \cite{Chen:2014}. Unlike previous proposal \cite{Peter:2014}, where the use of CZ gate was an integral part of cluster state creation, here the cluster state creation follows a macronode approach \cite{Flammia:2009}, which is entangled into a bilayer square lattice (BSL), comprising of two qumodes per layer. 
In the hybrid proposal \cite{Alexander:2016}, the frequency domain quantum wires are exposed to temporal delays and beam splitting, which results in cluster states having square lattice configuration in both time and frequency domains. This square lattice of temporal beam splitter is applied to every other frequency modes. A properly unbalanced Mach–Zehnder interferometer can easily separate the even and odd frequencies in quantum domain \cite{Huntington:2005}. In addition to the proposal, CV-based quantum computation is studied in detail to prove that such states are universal resources for universal MBQC \cite{Huntington:2005}.  

Although proposals for large scale CV cluster states have already been demonstrated, especially in time domain multiplexing as discussed in Section \ref{CV_TDM}, most of them are for 1D cluster states, which are not sufficient for universal 1WQC. Recently, 2D CV cluster states multiplexed in time domain are also proposed \cite{Larsen:2019,Asavanant:2019}, but are limited by the number of accessible qumodes in one of the dimensions (not uniform dimensions). Furthermore, extending to two dimensions introduces additional losses, which affects scalability beyond 1D in time domain multiplexing. Hybrid approaches (multiplexed in both temporal and frequency approaches), as discussed above, increase the size of smaller dimension significantly, but obtaining the phase reference for the simultaneous access of all frequency modes is still an open research problem. 

Keeping in mind the issues above, recently, in 2020, a hybrid approach exploiting the third-order Kerr non-linearity multiplexed in both time and frequency domain to create reconfigurable cluster states of one, two and three dimensions, is proposed in \cite{Wu:2020}. Time domain multiplexing provides sequential access for unlimited number of modes, whereas frequency domain multiplexing allows a simultaneous access to hundreds of highly connected modes. Obtaining phase references to simultaneously access all the frequency modes, which was a key challenge in the previous proposal, can be resolved by the third-order Kerr nonlinearity since it can create a frequency comb soliton, which can then act as phase reference, providing a simultaneous access to large number of frequency modes and hence can be scaled to 3D structure without introducing any additional losses. A microring resonator (MR) is used for generating classical frequency comb reference and eventually large-scale cluster states. Four wave mixing (FWM) approach is used for creating the side bands. The process of creating the cluster states of different dimensions includes sending a continuous wave pump field having a power below than parametric oscillation threshold. The resonator is pumped at even frequencies whereas the output field modes are detected only at odd frequencies. Pair-wise sidebands are created by coupling the FWM process with different cavity frequency modes. In \cite{Wu:2020}, TMS states (0D cluster states) are created, which are then combined by 50:50 beam splitter to produce 1D cluster states or dual-rail quantum wire. 

For the generation of 2D CV cluster state in \cite{Wu:2020}, the 1D cluster state is extended by adding an unbalanced Mach-Zehnder interferometer (UMZI), a 50:50 integrated beam splitter and a delay line. This approach is quite similar to that one proposed in \cite{Menicucci:2006}, for a 2D cluster state creation in time-frequency domain. However, a prominent advantage of this approach is an easier and more convenient integration on a photonic chip because of the shorter delay line, due to the larger bandwidth of MR. 


For 3D CV cluster states generation in \cite{Wu:2020}, the 2D cluster state setup were used with some modifications. Two copies of 2D cluster state generation setup were used but here balanced Mach-Zehnder interferometers (BMZIs) replaces the 50:50 integrated beam splitters, primarily because they can be tuned to work like 50:50 IBS and in such a way that CV cluster states can be created on the same chip.



\begin{figure}[!htp]
	\centering
		\includegraphics[scale=0.4]{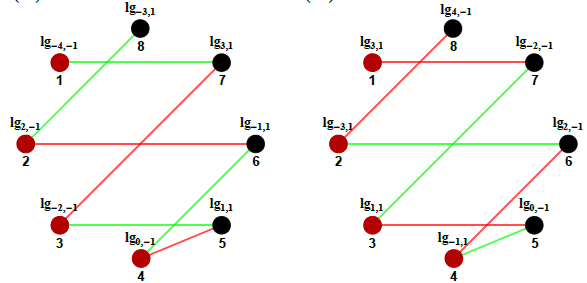}
	\caption{ Two possible graph representations of an eight-mode cluster state \cite{Yang:2020}. Black and blue lines indicate two different pumps)} 
	\label{fig:H7}
\end{figure}

Another hybrid approach for creating multipartite entangled states, this time using both frequency and spatial modes of an OPO is proposed recently in \cite{Yang:2020}. The proposed scheme can generate several cluster states in parallel. Moreover, the effect of finite squeezing on measurement of weighted graph states (graph states in CV domain can have weighted edges unlike qubit cluster states) is also discussed along with an illustration of the proposed approach for cluster states of 8 (Fig. \ref{fig:H7}), and 60 qumodes. In the experimental setup, two Laguerre-Gaussian (LG) pumps are used to pump the optical parametric amplification in a single OPO resulting in parallel creation of entangled modes in both spatial and frequency domains. 

All the state-of-the-art hybrid approaches for cluster states realization are summarized in Table \ref{tab:summary_CV-Domain}.

\begin{table*}[h]
    \centering
    \caption{Summary of state-of-the-art cluster states in CV domain}
    \begin{tabular}{p{0.8cm}|p{2.1cm}|p{2.1cm}|p{2.1cm}|p{2.2cm}|p{2.5cm}}
    \toprule
    Ref\# & Frequency Multiplexing & Time Multiplexing & Spatial Multiplexing &Cluster state dimension & No. of entangled modes\\
    \midrule
    \cite{Pysher:2011} &\hfil \Checkmark &\hfil-- &\hfil-- &\hfil 2D &\hfil$60^*$\\
    \hline
    \cite{Chen:2014} &\hfil \Checkmark &\hfil-- &\hfil-- &\hfil1D-dual rail &\hfil60\\
    \hline 
     \cite{WangPei:2014} &\hfil \Checkmark &\hfil-- &\hfil-- &\hfil1D,2D,3D &\hfil $10^4$\\
    \hline 
    \cite{zhu:2020} &\hfil \Checkmark &\hfil-- &\hfil-- &\hfil1D,2D,3D,nD &\hfil $10^4$\\
    \hline
    \cite{Yokoyama:2013} &\hfil -- &\hfil \Checkmark &\hfil-- &\hfil1D-dual rail &\hfil $10,000$\\
    \hline
    \cite{Yoshikawa:2016} &\hfil -- &\hfil \Checkmark &\hfil-- &\hfil1D-dual rail &\hfil $1.2\times10^6$\\
    \hline
    \cite{Larsen:2019} &\hfil -- &\hfil \Checkmark &\hfil-- &\hfil 2D &\hfil $30,000$\\
    \hline
    \cite{Asavanant:2019} &\hfil -- &\hfil \Checkmark &\hfil-- &\hfil 2D &\hfil $25000$\\
    \hline
    \cite{Pooser:2014} &\hfil -- &\hfil -- &\hfil\Checkmark &\hfil 1D-dual rail &\hfil $\simeq200$\\
    \hline
    \cite{yang:2016} &\hfil -- &\hfil -- &\hfil\Checkmark &\hfil 1D &\hfil 11-partite\\
    \hline
    \cite{Zhang:2017} &\hfil -- &\hfil -- &\hfil\Checkmark &\hfil 1D &\hfil $>$20\\
    \hline
     \cite{Cai:2018} &\hfil -- &\hfil -- &\hfil\Checkmark &\hfil 2D &\hfil 4\\
    \hline
    \cite{Peter:2014} &\hfil \Checkmark &\hfil \Checkmark &\hfil -- &\hfil 2D &\hfil --\\
    \hline
    \cite{Alexander:2016} &\hfil \Checkmark &\hfil \Checkmark &\hfil -- &\hfil 2D &\hfil $3\times10^3 \times \infty ^{**}$\\
    \hline
     \cite{Wu:2020} &\hfil \Checkmark &\hfil \Checkmark &\hfil -- &\hfil 3D &\hfil $>$2000\\
    \hline
    
    \cite{Yang:2020} &\hfil \Checkmark &\hfil -- &\hfil \Checkmark &\hfil -- &\hfil $60^\dagger$\\
    \bottomrule

\multicolumn{6}{l}{\footnotesize *15 copies of $2\times2$ cluster states}\\
\multicolumn{6}{l}{\footnotesize **$3\times10^3$ modes in frequency domain and infinite modes in time domain}\\
\multicolumn{6}{l}{\footnotesize $^\dagger$FSR between 1-10GHz can generate up to $10^3 - 10^4$ modes}\\

    \end{tabular}

    \label{tab:summary_CV-Domain}
\end{table*}


\section{DV-based Quantum Computation}\label{DVQC}
DV-based computation is the standard and original approach to perform quantum computation using discrete observables rather than continuous ones; that is, qubits rather than qumodes. 
More details on DV based quantum computation, particularly in comparison to CV based quantum computation, can be found in \cite{andersen:2015}. 

\subsection{Experimental Realization of DV Cluster States}\label{DV_exp_realization}
Creating cluster states entails carefully entangling the underlying qubits.
The cluster states in DV can be experimentally prepared by two methods: cooling the nearest-neighbor Ising-type Hamiltonian systems to its ground state or by dynamic implementation of CZ gates on a qubit lattice which are initialized in a superposition state $\ket{+}$, but measuring entanglement is a non-trivial. Since the 1989 discussion on entanglement of various mixed states by Werner \cite{Werner:1989}, many criteria for entanglement measurement have been proposed. However, symmetric extension criteria \cite{Doherty:2002} and partial transpose criteria \cite{Horodecki:1996,Peres:1996,Horodecki:2009} are the most widely used ones.

The photonic realization of cluster states focuses on the use of photons as potential qubits in a large cluster state, and are probabilistic as the cluster generation happens only upon the successful creation and detection of photons. Moreover, the quantum computation based on cluster states created by photons prepares the usual input states in $\ket{+}$ as part of initial cluster states. These attributes of DV-based cluster state generation limits their use whenever there is a need of deterministic application of unitary gates on an input state which is not prepared in the cluster state (output of previous computation) \cite{Ukai:2011} .
A number of research efforts have been made for realization of photonic cluster state enabling MBQC in photonic qubits. This section will summarize the proposals on photonic cluster states in MBQC context.

In \cite{Walther:2005}, four photons were polarized to realize a four-qubit cluster state. Quantum state tomography \cite{Ariano:2002} was used for the extraction of density matrix of a quantum state. The cluster state fidelity of $0.60\pm0.02$ was achieved surpassing the local realism threshold of 0.56. Arbitrary single qubit rotations SU(2) and two qubit operations (CPhase and CNOT) were implemented with an average fidelity rate of $0.85\pm0.04$. and $0.93\pm0.01$ respectively. Grover’s search algorithm \cite{Grover:1996} with success probability of 90\% was also implemented.   

Another proof-of-principle demonstration of 1WQC was presented in \cite{Chen:2007}, where a two-photon four-qubit cluster state resource is developed and is entangled in both spatial and polarization modes of photons, with a state fidelity rate exceeding 88\%. Afterwards, two-qubit quantum gates were implemented with around 95\% average fidelity rates and grover algorithm with success probability of around 96\%. 

In \cite{Vallone:2008}, a four-qubit cluster state with fidelity rate of  $0.880\pm0.013$ was realized by exploiting the full entanglement of two photons having two degrees of freedom (linear momentum and polarization). Arbitrary single qubit rotations and two qubit operations (CNOT and CPhase) gates were implemented with average fidelity rates of $0.867\pm0.018$ and $0.907\pm0.010$ respectively.  

Hyperentanglement \cite{Barreiro:2005} has been used widely to increase the number of qubits without increasing the photons. In this approach, the particles (photons) are entangled in various degrees of freedom \cite{Vallone:2007,Chen:2007,Lanyon:2009,Gao:2010a}, which can potentially increase fidelity and qubit generation rate. Since 1WQC requires high number of qubits to create cluster state \cite{Gao:2010} hyperentanglement is very well suited for it's physical realization. Based on same approach, a two-photon six-qubit linear cluster state was developed in \cite{Ceccarelli:2009}, where each particle encodes three qubits exploiting two distinct degrees of freedom of photons. Linear momentum encodes two qubits whereas a single qubit is encoded in photon polaritztion. The cluster state fidelity of $0.6350\pm0.0008$ was achieved, better than the previously proposed six-qubit cluster states. In a similar fashion, four photons were used to develop a six-qubit cluster state in \cite{Gao:2010}. The photons were entangled in polarization and spatial modes. The overall cluster state fidelity achieved was $0.61\pm0.01$\% and a CNOT gate with fidelity rate of $79\pm1$\% was also implemented using the created cluster state. 

In \cite{Tokunaga:2008}, four photons in horizontal polarization were used to develop a four-qubit cluster state. The average cluster state fidelity of $0.860\pm0.015$ was achieved. Furthermore, arbitrary single qubit rotations were also implemented exploiting the created cluster state with an average fidelity rate of $0.926\pm0.10$.
The fusion of path qubits is also one of the techniques to create cluster states with larger number of qubits with relatively lesser particles. In \cite{Lee:2012}, a seven-qubit entangled cluster state was created by fusing two distinct two-photon four qubit linear graph states with an average fidelity rate of $>$64\%. To demonstrate the authenticity of developed cluster state two-bit Deutsch-Jozsa \cite{Deutsch:1992} algorithm was implemented with success probability of $>$90\%.

In \cite{Bell:2013}, a high-brightness photonic crystal fiber (PCF) is used as an entangled photon pairs source \cite{Fulconis:2007,clark:2011} to create a four-photon star state is formed from these photon pairs by using a post selected fusion gate. A scalable formation of larger cluster states can be achieved using fusion gates. The term scalable here refers to begin with smaller entangled states source (Bell states) \cite{Browne:2005,Bodiya:2006} and going up to larger states. The fused four-photon star cluster state has fidelity rate of $0.66\pm0.01$ was achieved. Single-qubit (Hadamard) gate was implemented with fidelity rate of $0.67\pm0.03$. Moroever, two-qubit (CNOT) gate  and three-qubit (Toffoli) gate were implemented with fidelity rate of $0.64\pm0.01$ and $0.76\pm0.04$ respectively. The state-of-the-art photonic cluster states realization are summarized in Table \ref{tab:DV_table_1}. 

The core of cluster states is entanglement, which might be challenging to create and maintain. In the following sub-sections, some recent proposals on qubits entanglement primarily in photonic  and superconducting qubits, which potentially can be used as cluster state for MBQC, are discussed.

\begin{table*}[!htp]
    \centering
    \caption{State-of-the-art Cluster states in using discrete variables}
    \begin{tabular}{p{0.6cm}|p{1.8cm}|p{1.8cm}|p{1.9cm}|p{1.7cm}|p{1.7cm}|p{1.6cm}|p{1.3cm}}
    \toprule
         Ref\# &Generated Cluster state &Implemented Algorithm &Cluster state fidelity &\multicolumn{3}{c|}{Average gate fidelity} & Algorithm success \\
         \cline{5-7}
          &&& & Single qubit & Two qubit & Three qubit &\\
          \midrule
    \cite{Walther:2005} &four-photon four-qubit &Grover's search algorithm &$0.60\pm0.02$ &$0.85\pm0.04$ &$0.93\pm0.01$ &-- &0.90 \\
    \hline
    \cite{Chen:2007}  &Two-photon four-qubit &Grover's search algorithm &$>0.88$ &-- &$0.95$ &-- &0.96 \\
    
    \hline
    \cite{Lee:2012}  &Seven-qubit &Deutsch-Jozsa algorithm &$>0.64$ &-- &-- &-- &$>0.90$ \\
    
     \hline
    \cite{Vallone:2008}  &Two-photon four-qubit &-- &$0.880\pm0.013$ &$0.867\pm0.018$ &$0.907\pm0.010$ &-- &-- \\
    
    \hline
    \cite{Ceccarelli:2009}  &Two-photon six-qubit &-- &$0.635\pm0.0008$ &-- &-- &-- &-- \\
    
     \hline
    \cite{Gao:2010}  &Four-photon six-qubit &-- &$0.61\pm0.01$ &-- &$79\pm1$ &-- &-- \\
    
    \hline
    \cite{Tokunaga:2008}  &Four-photon Four-qubit &-- &$0.860\pm0.010$ &$0.926\pm0.015$ &-- &-- &-- \\
    
    \hline
    \cite{Bell:2013}  &Four-photon &-- &$0.66\pm0.01$ &$0.67\pm0.03$ &$0.64\pm0.01$ &$0.76\pm0.04$ &-- \\
    
    \bottomrule
    \end{tabular}
    
    \label{tab:DV_table_1}
\end{table*}

\subsubsection{Entanglement in Photonic Qubits}\label{photon_entanglement}

Photonic qubits have the potential to be the future of QIP because of the advantages they offer over other candidates, such as improved speed, larger bandwidth,  compatibility with CMOS fabrication, and most importantly improved decoherence, particularly for single photons. Hence, single photons are widely used for the realization of high precision quantum gates and cluster states realization on MBQC.

The first experimental demonstration of three spatially separated photons was proposed in \cite{Bouwmeester:1999}. Since then, multiple efforts have been made for entangling various number of qubits \cite{sackett:2000,Haffner:2005,Prevedel:2009,Monz:2011,Wieczorek:2009,raadmark:2009}, most of them are based on spontaneous parametric down conversion (SPDC) \cite{Kwiat:1995}.
In 2012, a total of eight individual photons were successfully entangled \cite{Yao:2012} using an ultra-bright sources of entangled photon pairs \cite{Kim:2003} along with an eight-photon interferometer and post-selection detection. A total of four photon pairs were used in \cite{Yao:2012} which were then, by appropriate experimental setup, transformed to eight-photon GHZ state, with state fidelity of about $0.708\pm0.016$, which is greater than 0.5 threshold assuring the genuine multi-partite entanglement \cite{Dur:2001}.


Later in 2016, an approach to entangle 10 spatially separated entangled single photons was proposed in \cite{Wang:2016}. The resulting state has the fidelity of about 0.57 greater than the genuine multi-partite entanglement threshold and hence confirming the genuine multi-partite entanglement between all the photonic qubits. The key factors in achieving 10-photon entanglement include SPDC photon pair source with high brightness, high collection efficiency and high photon indistinguishability. 


In 2018, an experimental demonstration of 12-photon entanglement was proposed \cite{Zhong:2018}, by using an optimal SPDC entangled photon source with high indisguishability (96\%), and efficiency (97\%). The resulting state fidelity of $0.572\pm0.024$ was achieved ensuring genuine multi-partite entanglement between all the entangled qubits. In the same year, an experimental demonstration of entangling 18-qubit GHZ state was also proposed in \cite{Wang:2018} by exploiting three distinct degree of freedoms (polarization, spatial modes, and orbital angular momentum) of six photons. A genuine multi-partite entanglement was confirmed between all 18 qubits with resulting state fidelity of $0.708 \pm0.016$ which quite higher than 12-photon entanglement \cite{Zhong:2018}. 

In all the proposals discussed above, probabilistic sources were used for entanglement creation which are restricted in terms of efficiency and high hardware cost. Recently, in 2020, a resource efficient sequential creation of linear cluster states from a single photon emitter was proposed \cite{Istrati:2020}. A single entangling gate in a fiber loop configuration \cite{Senellart:2017} was used for entangling the stream of incoming photonic qubits. A four-photon linear cluster state was generated; however the proposed setup is capable of creating a cluster state of arbitrary number of photons. Hence, the proposed approach exhibits tremendous scaling potential. The said proposal makes use of semiconductor quantum dots for single photon creation sequentially \cite{Pilnyak:2017}, and temporal delay loop approach for entanglement generation \cite{Senellart:2017}. The photon indistinguishability is measured at different time intervals and it ranges from 77\% to 95\%. Moreoevr, the genuine multi-partite entanglement has been verified in all four photons. 

All the entanglement advances using photonic qubits are summarized in Table \ref{tab:recent_breakthrough_DV}.

\subsubsection{Entanglement in Superconducting Qubits}\label{supercon_entanglement}
Superconducting circuits are another promising approach for the realization of qubits and qubit gates. The advantages like scalablity, ability of easy coupling and control and high designability of superconducting circuits make them attractive for the physical realization of quantum computers \cite{huang:2020}. Here, we report some recent proposals focusing on entangling the superconducting qubits which can potentially serve as resource states for MBQC. 

\begin{figure}[!htp]
	\centering
		\includegraphics[scale=0.3]{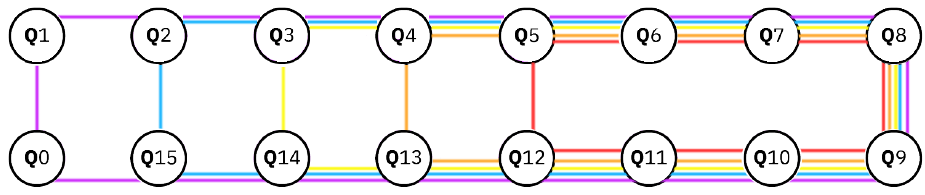}
	\caption{Graph states created in \cite{Wang_2018a}. Red lines indicate the 8-qubit graph state, orange lines indicate the 10-qubit graph state and 12, 14 and 16-qubit graph states are represented by yellow, blue and purple lines respectively} 
	\label{fig:DV4}
\end{figure}

In 2018, 16 superconducting qubits of IBM’s quantum device ibmqx5 (16-qubit quantum device) were entangled using optimized low-depth circuits \cite{Wang_2018a}. 
The graph states which are the generalization of cluster states \cite{Raussendorf:2001}, were used for entangling of the qubits. The starting point for graph state preparation was to build a circuit as per the definition of graph states (Section \ref{clusterstates}). For entanglement between qubits  CZ gates were implemented using one CNOT and two H-gates. which eventually produces  between the qubits. The total of 5 graph states having ring configuration were created as shown in Fig. \ref{fig:DV4}.

For entanglement detection in \cite{Wang_2018a} reduced density matrices-based entanglement criterion up to 4 qubits (using maximum likelihood method \cite{Smolin:2012}) was used and negativity of resulting two-qubit systems were calculated. All the negativity values for 8-qubit graph state were calculated to be significantly greater than zero and hence fully entangled. In case of 10-qubit graph state 9 out of 10 negativity values were greater than zero and is fully entangled, since it is claimed that for an $n$-qubit system if $n-1$ qubits are entangled then the whole system is entangled. For 12 and 14-qubit graph states, two of the negativities were calculated to be zero (not entangled), however with proper separation of qubits in these states and reduced density matrix calculations, it was proved that these states are also fully entangled. For a 16-qubit graph state 15 out of 16 negativities were greater than zero and hence fully entangled. 

\begin{figure}[!htp]
	\centering
		\begin{subfigure}[b]{0.3\textwidth}
         \centering
         \includegraphics[scale=0.28]{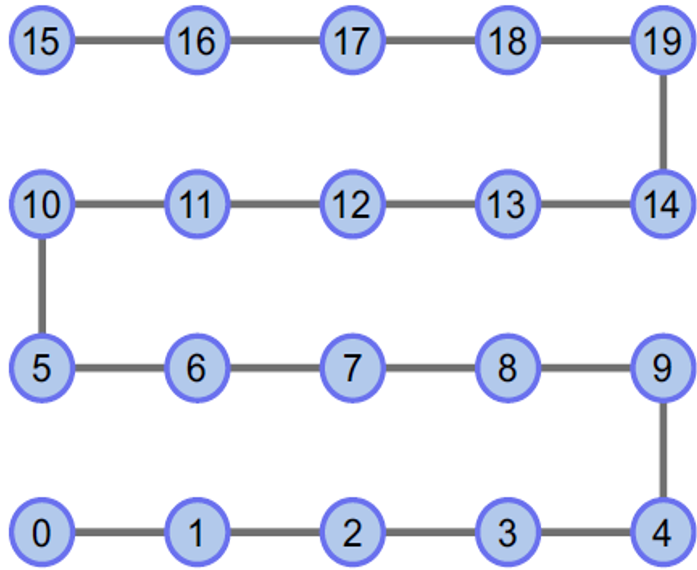}
         \caption{}
         \label{fig:DV5_a}
     \end{subfigure}
     \quad
     \hspace{2.5cm}
     \begin{subfigure}[b]{0.3\textwidth}
         \centering
         \includegraphics[scale=0.28]{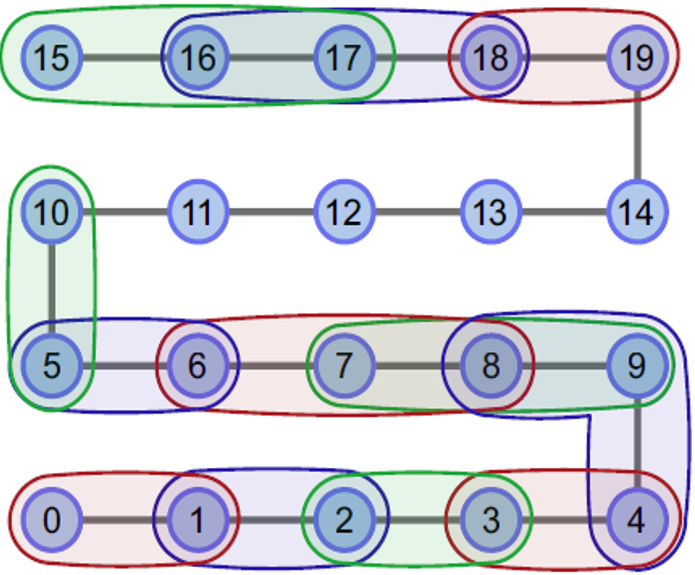}
         \caption{}
         \label{fig:DV5_b}
         \end{subfigure}
	\caption{(a) Embedded Graph state on Poughkeepsie device where nodes represents qubits and edges represents two-qubit operations to create entanglement (b) Generated 20-qubit graph state highlighting regions of genuine multi-partite entanglement. } 
	\label{fig:DV5}
\end{figure}

Another proposal entangling 20 qubits in a superconducting quantum computer was proposed recently in \cite{Mooney:2019}. The entanglement was created on a 20 superconducting qubit IBM device (Poughkeepsie), and it was shown that all its 20 qubits can be entangled. A Graph state was prepared along the path of all 20 qubits following almost the same methodology as in \cite{Wang_2018a}. the embedded graph state along with the entanglement between various qubits is shown in Fig. \ref{fig:DV5}. Afterwards, full quantum state tomography was performed on every four connected qubits forming a chain along the path and negativity of every qubit pair was evaluated for entanglement detection. It was shown that every qubit pair is entangled leading to a conclusion that whole graph state is entangled and has better magnitue than \cite{Wang_2018a}. Moreover, using entanglement witness it was shown that the chains of three qubits exhibits genuine multi-partite entanglement. 
  
Recently in 2019, a genuine 12-qubit entanglement in linear configuration has been prepared and verified in a superconducting processor \cite{Gong:2019}. In this proposal the CZ gates are applied for the entanglement creation between the qubits in the processor. The superconducting processor used in the experiment consists of 12 transmon qubits \cite{Koch:2007}, each having a fast flux-bias line (Z), a microwave drive line (XY) and a readout resonator. For the entanglement creation process, the qubits are initialized in state $\ket{0}$ and relaxed for 300$\mu$s. A total of 11 controlled phase (CZ) gates were used for entangling 12 qubits with fidelity of $0.5544\pm0.0025$.



\begin{figure}[!htp]
	\centering
		\begin{subfigure}[b]{0.3\textwidth}
         \centering
         \includegraphics[scale=0.4]{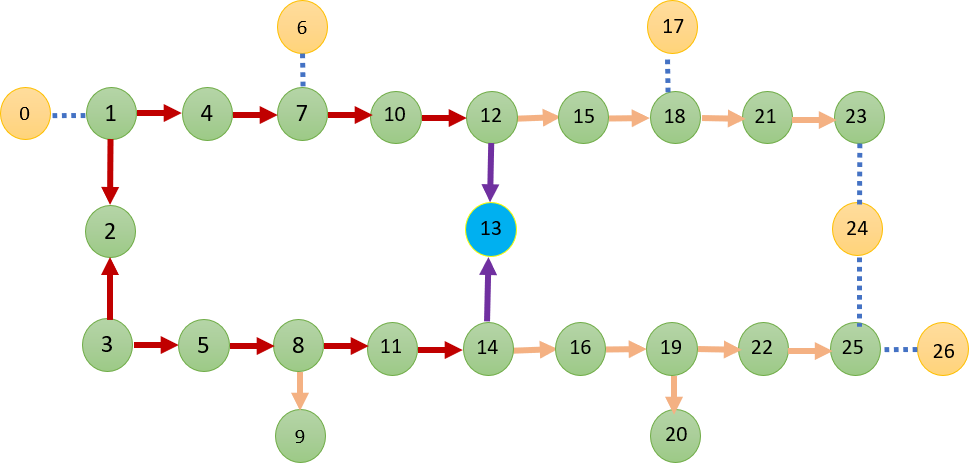}
         \caption{}
         \label{fig:DV7_a}
     \end{subfigure}
     \qquad
     \hspace{3cm}
     \begin{subfigure}[b]{0.4\textwidth}
         \centering
         \includegraphics[scale=0.4]{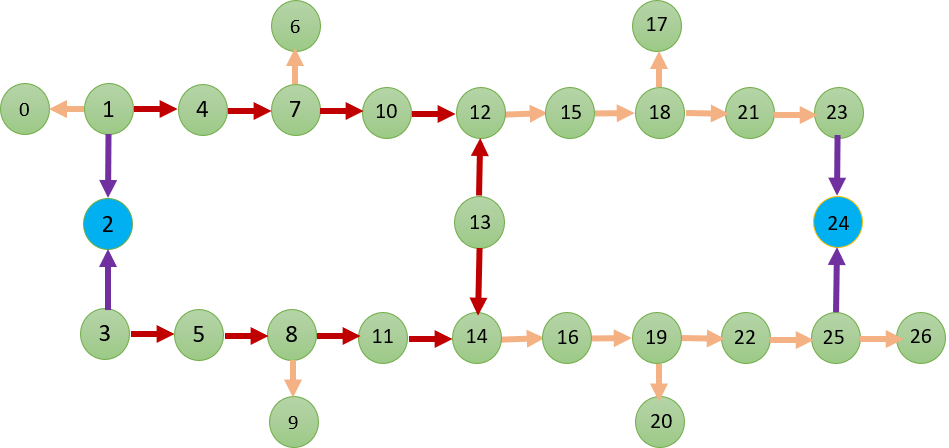}
         \caption{}
         \label{fig:DV7_b}
         \end{subfigure}
	\caption{Created GHZ states on ibmq\_montreal device \cite{mooney:2021}. The light green circles indicate qubits whereas light blue circles indicate the parity check qubits. The qubits not directly contributing to the experiment are represented by light orange circles.  The arrows indicate the CNOT operation as: control\_qubit$\rightarrow$ target\_qubit. The CNOT operations directly involved in GHZ state creation are represented by dark brown arrows. Light brown arrows represent CNOTs increasing the state size, purple arrows represent the CNOT operation involved in parity check. No two-qubit operation is applied on the qubits connected by dotted blue line } 
	\label{fig:DV7}
\end{figure}

In 2021, 27 superconducting qubits having the potential of being used as central resource for MBQC is proposed in \cite{mooney:2021}. The experiment was performed on ibmq\_montreal device which comprises of 27 superconducting transmon qubits \cite{Koch:2007}. Alongside the entanglement, quantum read out error mitigation is also implemented \cite{Maciejewski:2020}, since the faulty measurements can affect the entanglement strength. Parity verification-a basic protocol for quantum error correction, was also investigated on resulting state fidelity. A total of two GHZ states (11-qubit state extended up to 19 qubits and 19-qubit state extended up to 25 qubits), were created with one and two parity check qubits respectively and are shown in Fig. \ref{fig:DV7}. 
Additionally 26 and 27-qubit states were created with fidelity rates of $0.62\pm0.06$ and $0.61\pm0.05$ using quantum read out measurement, with genuine multi-partite entanglement across all the qubits.

Right after the 27-qubit entanglement on ibmq\_montreal device \cite{mooney:2021}, a largest till date, 53-qubit and 65-qubit IBM devices namely ibmq\_rochester and ibmq\_manhattan respectively, were entangled along with quantum read out measurements \cite{mooney:2021a}. On 53-qubit ibmq\_rochester device, without applying the quantum read out measurement technique resulting in entanglement of 31 out of 58 qubit pairs were entangled without applying the quantum read out measurement technique . The largest entangled region has maximum of 9 qubits. Similarly, 56 out of 58 qubit pairs were entangled with quantum read out measurement applied. The largest entangled region consists of all qubits of the device. 

         

On ibmq\_manhattan device, all the 72 qubit pairs were entangled in both cases (without applying the quantum read out measurement and with quantum read out measurement applied). 
Hence all 65 qubits of the device were successfully entangled. This is the largest qubit entanglement reported till date. The recent breakthroughs in qubit entanglement, which can potentially lead towards scalable MBQC are summarized in Table \ref{tab:recent_breakthrough_DV}.

         

\begin{table*}[!htp]
    \centering
    \caption{Recent Breakthroughs in Qubits Entanglement}
    \begin{tabular}{p{0.6cm}|p{2.7cm}|p{3cm}|p{3cm}|p{3cm}}
    \toprule
         Ref \# &Qubits type &No. of entangled qubits &Genuinely multi-partite entangled qubits &Resulting state fidelity\\
     \midrule
     \cite{Yao:2012} &\hfil Photonic &\hfil 8 &\hfil all &\hfil$0.708\pm0.016$\\
     \hline
     \cite{Wang:2016} &\hfil Photonic &\hfil 10 &\hfil all &\hfil$0.573$\\
     \hline
     \cite{Zhong:2018} &\hfil Photonic &\hfil 12 &\hfil all &\hfil$0.708\pm0.016$\\
     \hline
     \cite{Wang:2018} &\hfil Photonic &\hfil 18 &\hfil all &\hfil $0.572\pm0.024$\\
     \hline
     \cite{Istrati:2020} &\hfil Photonic &\hfil $4^*$ &\hfil all &\hfil --\\
     \hline
     \cite{Wei:2020} &\hfil Superconducting &\hfil 20 &\hfil 18 &\hfil $0.5165\pm0.0036$\\
     \hline
     \cite{Gong:2019} &\hfil Superconducting &\hfil 12 &\hfil all &\hfil $0.5544\pm0.0025$\\
     \hline
    \cite{Mooney:2019} &\hfil Superconducting &\hfil 20 &\hfil all &\hfil --\\
         \hline
    \cite{Wang_2018a} &\hfil Superconducting &\hfil 16 &\hfil all &\hfil --\\
         \hline
     \cite{mooney:2021} &\hfil Superconducting &\hfil \makecell{26\\27} &\hfil all &\hfil \makecell{$0.62\pm0.06$\\$0.61\pm0.05$}\\
         \hline
    \cite{mooney:2021a} &\hfil Superconducting &\hfil \makecell{53 \\ 65} &\hfil all &\hfil --\\
    \bottomrule
    
\multicolumn{5}{l}{\footnotesize * The proposed setup is capable to generate linear cluster having arbitrary number of photonic qubits}
    \end{tabular}
    
    \label{tab:recent_breakthrough_DV}
\end{table*}

\section{Conclusion and Outlook}\label{conclusion}
The race to achieve universal scalable quantum computation paradigm is in full swing over the last decade. MBQC is one of the leading candidates exhibiting promising potential for scalable quantum computation. The computations in 1WQC completely 
relies on a highly entangled resource state (cluster state) and hence cluster states are integral part of 1WQC. Generation of such state, typically with higher dimensions (2D), already achieves universal quantum computation. Going beyond 2D, cluster states can potentially even achieve fault-tolerance. Over the last few years, development of these cluster states has attracted the research community, with two different main approaches: CV cluster states and DV cluster states. A CV cluster state is the combination of large number of entangled qumodes arranged in a cluster, whereas DV cluster state is the combination large number of qubits.

In this paper, we provide a comprehensive compilation and comparison of the techniques and approaches being used for cluster states realization in both CV and DV. While similar surveys exist \cite{Pfister:2019,Tzu:2021}, these were not focused on the physical realization of MBQC.  


CV cluster states can be realized on different approaches, namely: FDM, TDM, SDM and hybird. Among those, FDM and TDM are the most widely used ones. 
While FDM can generate cluster states with more than 3D dimensions, TDM can only generate 3D cluster states. 
Additionally, TDM can generate infinite number of qumodes, but allows only sequential access to qumodes, whereas FDM can generate sufficiently large number of qumodes while allowing simultaneous access them, which gives it slight edge over TDM. 
The SDM approach has also been explored with the largest cluster state dimension of 2D and limited number qumodes but still having the potential to scale up to large number
of qubits and increased cluster state dimension. Recently, hybrid proposals for CV cluster state realization have been proposed using a combination of 
two multiplexing approaches. The largest hybrid CV cluster state of dimension up to 3D exploiting FDM and TDM has been reported in the literature. All the CV approaches are still in the developing and are equally likely to dominate in the future. 

DV cluster states are based on qubits, and have been physically realized following different approaches, most notably: photons and superconducting circuits, where significant amount of research has been conducted already. These qubits can then be used to form the cluster states as required by MBQC. 
On photonic, large number photonic qubits have been successfully entangled, and research in this approach already looks promising. On superconducting, even larger number of qubits have been successful entangled, and research is progressing rapidly in this area.

Based on the surveyed literature, it seems that the CV track is currently leading the way towards scalable MBQC with cluster states dimensions of beyond 3D with thousands of entangled qumodes, whereas the largest number of entangled qubits reported in literature is 65. This smaller qubit number limitation is primarily due to the complex experimental procedure for qubits preparation and entanglement generation between them in DV. 
In summary, both CV and DV cluster state realizations are independently progressing at a rapid pace and it is hard to predict which approach could potentially be the future of scalable universal MBQC. The answer will not be straightforward, and may not even be either one, but (possibly) a combination. 

Recently, entanglement generation by bridging both CV and DV approaches have been proposed in single hybrid experiments \cite{andersen:2015}. For example, an experimental illustration of teleporting DV qubits by using CV techniques has been presented in \cite{Takeda:2013}. Entanglement generation between CV and DV qubits in \cite{Ho:2014,jeong:2014} provided a foundation to communicate between DV and CV nodes and towards the realization of hybrid quantum networks.  
MBQC can also be benefit from such hybrid approach where DV (non-Gaussian) projectors can be used to perform computations exploiting CV (Gaussian) cluster states \cite{andersen:2015}. Moreover, deterministic generation of multi-qubit gates in DV (particularly photonic) using MBQC is challenging as compared to CV and requires complex experimental setups. A combination of CV homodyne measurement and a relatively weak cross-Kerr nonlinearity can realize quantum non-demolition measurement making it possible to almost deterministically  generate multi-qubit gate in DV with minimal experimental resources, which otherwise is extremely difficult \cite{Nemoto:2004}.  
To the best of our knowledge such approach is yet to be explored in the literature, adding one more direction for a potential fault-tolerant quantum computer based on MBQC.

\label{Bibliography}

\bibliographystyle{unsrt} 
\bibliography{main}
\end{document}